\documentclass[aps,notitlepage,twocolumn,nofootinbib,pra,10pt]{revtex4-2}
\usepackage{amsmath,amsfonts,amssymb,amsthm,bbm,graphicx,enumerate,times,mathtools,hyperref,physics,marvosym,float,fancyref,etoolbox,appendix,soul,enumitem,esint,dsfont,bm,subfigure}
\hypersetup{colorlinks,linkcolor=blue,citecolor=red,urlcolor=blue}
\usepackage[ruled,longend]{algorithm2e}
\usepackage[dvipsnames]{xcolor}
\usepackage[noabbrev,capitalise]{cleveref}
\usepackage[mathscr]{euscript}

\creflabelformat{equation}{#2#1#3}
\usepackage[ruled,longend]{algorithm2e}
\SetKwComment{Comment}{/* }{ */}
\newcommand{\Assert}[2]{\textbf{Assert:} #1 \newline}

\DeclareMathOperator*{\argmin}{arg\,min}

\begin{document}
\title{Lightcone Bounds for Quantum Circuit Mapping via Uncomplexity}
\author{Matthew Steinberg$^{1,2}$*}
\author{Medina Bandi{\'c}$^{1,2}$*}
\author{Sacha Szkudlarek$^{1}$}
\author{Carmen G. Almudever$^{3}$}
\author{Aritra Sarkar$^{1,2}$}
\author{Sebastian Feld$^{1,2}$}
\affiliation{$^{1}$Quantum Computing Division, QuTech, Delft University of Technology, Delft, the Netherlands}
\affiliation{$^{2}$Department of Quantum \& Computer Engineering, Delft University of Technology, Delft, the Netherlands}
\affiliation{$^{3}$Computer Engineering Department, Technical University of Valencia, Valencia, Spain}
\date{\today}

\begin{abstract}
Efficiently mapping quantum circuits onto hardware is an integral part of the quantum compilation process, wherein a circuit is modified in accordance with the stringent architectural demands of a quantum processor. Many techniques exist for solving the quantum circuit mapping problem, in addition to several theoretical perspectives that relate quantum circuit mapping to problems in classical computer science. This work considers a novel perspective on quantum circuit mapping, in which the routing process of a simplified circuit is viewed as a composition of quantum operations acting on density matrices representing the quantum circuit and processor. Drawing on insight from recent advances in quantum circuit complexity and information geometry, we show that a minimal SWAP-gate count for executing a quantum circuit on a device emerges via the minimization of the distance between quantum states using the quantum Jensen-Shannon divergence, which we dub the lightcone bound. Additionally, we develop a novel initial placement algorithm based on a graph similarity search that selects the partition nearest to a graph isomorphism between interaction and coupling graphs. From these two ingredients, we construct an algorithm for calculating the lightcone bound, which is directly compared alongside the IBM Qiskit compiler for over $600$ realistic benchmark experiments, as well as against a brute-force method for smaller benchmarks. In our simulations, we unambiguously find that neither the brute-force method nor the Qiskit compiler surpasses our bound, signaling utility for estimating minimal overhead when realizing quantum algorithms on constrained quantum hardware. This work also constitutes the first use of quantum circuit uncomplexity to practically-relevant quantum computing. We anticipate that this method may have diverse applicability outside of the scope of quantum information science. 
\end{abstract}

\maketitle

\def\thefootnote{*}\footnotetext{These authors contributed equally to this work. The corresponding author can be reached at \url{m.a.steinberg@tudelft.nl}.}\def\thefootnote{\arabic{footnote}}


\section{Introduction}

The promise of quantum technology extends to many areas of modern theoretical physics, computer science and cryptography, among others \cite{nielsenchuang}. In spite of much success over the past 30 years, current-generation quantum technology is characterized by noisy, intermediate-scale devices that are severely limited not only by the depth and size of the quantum circuits that can be executed, but also by the qubit connectivity of such devices \cite{preskill2018quantum}. Such processors have allowed for the first generation of quantum-technology demonstrations, ranging from experimental realizations of hybrid quantum-classical optimization techniques \cite{classquant-trappedion,firstelectronicstructure} to resource-intensive algorithms such as fault-tolerant quantum error-correction codes (QECCs) \cite{1stQEC,realizationQEC,terhal2015quantum}.  

With such promise as is forecasted for quantum technology, \emph{full-stack} design approaches have emerged, in order to delegate resources efficiently and to ensure high success rates for a given quantum circuit, realized on a quantum processor \cite{murali2019full,bandic2022full, Bertels_fullstack}. As such, one of the cardinal issues to emerge for practical quantum computing is that of \emph{quantum compilation}, which can be broadly defined as the various engineering-level steps required to translate and prepare a quantum circuit for execution on a quantum processor \cite{bandic2020structured}. Central to quantum compilation is the \emph{quantum circuit-mapping problem} (QCMP), which concerns the assignment and rearrangement of qubits from an algorithm to a processor as a quantum circuit is executed, in order to guarantee high fidelity of the resulting state \cite{venturelli_tackling_qubit_mapping,qubitallocation}. It is known that the quantum circuit-mapping problem is NP-complete \cite{OnQubitMappingProblem,alon1993routing}, and has been likened to the traveling salesman problem (TSP) on a torus \cite{paler2021nisqtorus}. The QCMP is also related to token swapping \cite{miltzow2016approximationtokenswap}. 
Many competing approaches have been proposed for solving the QCMP, with all of the state-of-the-art strategies trading accuracy for speed, among other considerations \cite{wagner2023improving,  murali2019noise, tannu2019not, li2020towards, zulehner2018efficient, venturelli2019quantum, lao2018mapping, lao2019mapping, herbert2018using,lye2015determining, li2020qubit, biuki2022exact, molavi2022qubit, moro2021quantum, devulapalli2022quantum, upadhyay2022shuttle,nottingham2023decomposing}. However, to our knowledge, no work has attempted to formulate the QCMP from a standpoint grounded in theoretical physics and quantum information theory. The motivation for such an endeavor is twofold. Firstly, since the QCMP is a physical process, such a description can provide new insights and perspectives on how best to solve it. Secondly, by providing a fundamental description of the QCMP, we lay the groundwork for uniting various contemporary approaches towards a solution, and show how they compare to each other in a self-consistent framework. In short, a physics-motivated description of the QCMP offers consensus for current and future solution strategies, and how best to compare them. 

Underpinning the advances in quantum technology, \emph{quantum information theory} seeks to quantify the achievable limits of information processing on a fundamental mathematical basis using quantum physics \cite{nielsenchuang,wilde_qit,watrous2018theory}. While much progress is already notable, many fields outside of the immediate scope of quantum information theory have benefited from incorporating quantum-information-theoretic interpretations to outstanding research problems, including theoretical physics \cite{quantumHEPth,jerusalemBlackHoles}, network science (which studies the behavior of complex networks from the standpoint of statistical mechanics and graph theory) \cite{biamonte_spectral,biamonte_nature,severini}, among many others. Bearing in mind such potential, we apply the machinery of quantum information theory, in particular, \emph{quantum circuit complexity} \cite{susskind2ndlawcomplexity,yungerhalpern_uncomplexity}, \emph{spectral graph entropy} \cite{biamonte_spectral,asymptoticentropy_diffusiontimepaper,properties_qJSD,metriccharacter_qJSD}, and the \emph{quantum operations} formalism \cite{wolf_qchannels,wilde_qit,Christandl_qit_lectures,nielsen_majorization_book} to the QCMP, in order to describe the problem of preparing certain quantum states on a quantum processor whose qubit connectivity is restricted. As a quantum circuit itself describes a sequence of unitary transformations under which a quantum state transforms, addressing such quantum operations under the guise of a processor's connectivity is not only reasonable using quantum information theory, but also embodies a natural extension of quantum information theory to the setting of practical quantum computing.

Several recent proposals have sought to establish links between graph theory and the QCMP \cite{steinberg1,siraichi2018qubit,siraichi2019qubit,li2020qubit,biuki2022exact}. Since then, it has become commonplace in the literature to consider an \emph{interaction graph} (IG), in which edges represent the necessary two-qubit interactions for implementing a quantum circuit, and a \emph{coupling graph} (CG), whose edges determine allowed two-qubit interactions between neighboring subsystems on the processor \cite{bandic2023interaction}. In many of these proposals, the SWAP-gate count required to realize a quantum algorithm on a given quantum processor is considered to be a typical metric for the objective function of a mapping strategy \cite{zulehner2018efficient}. 

In this vein, we strengthen this connection by initiating a study of the QCMP from the theoretical standpoint, using graph theory and network science as a foundation. More concretely, we propose a special case of the QCMP in which all two-qubit interactions of a given IG can be compressed into a single time slice of the quantum circuit; this simplification can be likened to a sort of ``lightcone" path through a configuration space, which we explain in detail in \cref{section:discussion}. Starting from this point, we translate the IG and CG into density matrices, and calculate their \emph{thermodynamic path length} in the configuration space of density matrices, given certain allowed superoperator transformations. These allowed superoperator transformations consist of a combination of \emph{doubly-stochastic quantum operations} \cite{nielsen_majorization_book,sagawa_majorization_book} to permute vertices of the CG, and Bell measurements on the IG, in order to sequentially and methodically minimize the path length over the configuration-space geodesic. Using recent results in \emph{quantum circuit complexity} \cite{yungerhalpern_uncomplexity,susskind2ndlawcomplexity} and methods from \emph{quantum information geometry} \cite{bengtsson2017geometry_of_quantum_states_book,datta2020relating,liu2020quantumFI}, we carefully show that entropic divergence measures can be used in order to minimize the distance between density matrices describing the IG and CG along the configuration-space geodesic of allowed quantum operations, and that a minimal SWAP-gate count can be ascertained using this method. This minimal SWAP-gate count is shown to coincide with the \emph{quantum circuit uncomplexity} \cite{yungerhalpern_uncomplexity,susskind2ndlawcomplexity}; as such, we name this lower bound the \emph{SWAP uncomplexity}. As this lower bound does not take into account the traditionally-used gate-dependency graph of the IG, the SWAP uncomplexity represents a lightcone solution to the QCMP, in which infinite parallelization of two-qubit gates is possible. 

In addition, we develop a novel algorithm for the \emph{qubit assignment} (or initial placement) of qubits from the IG to the CG, based on a subgraph isomorphism and graph similarity search \cite{zager2008graph,koutra2011algorithms, samanvi2015subgraph}, which has applications for \emph{multi-programming} on a quantum device \cite{niu2022multi} and may be of independent interest. In this case, however, it serves as a necessary step in our formulation and further enables our approach by constraining the coupling graph to match the size of an IG, which is one of the method's crucial requirements. This algorithm also facilitates a calculation that we use to compute a maximal SWAP-gate count.

Together with the formalism introduced, a combined approach is constructed that searches for the best qubit assignment in terms of the \emph{graph-edit distance} (GED), and then calculates the SWAP uncomplexity, given an IG/CG pair as inputs. We test the resultant algorithm against IBM's Qiskit compiler, finding that in all cases, the SWAP-gate count as calculated by the SWAP uncomplexity algorithm is never surpassed, in full agreement with our formalism. This lower bound is of great importance, as such constraints can help with the prediction of compilation performance, as well as for making design choices relevant in application-specific mapping strategies and quantum devices \cite{murali2019full}.

This article is organized as follows. We review the QCMP in \cref{section:QCMP_Background}. Following this section, we immediately discuss in \cref{section:graphs_as_density_matrices} how to mathematically map the \emph{graph Laplacian} of a simple graph into the density matrix formalism, while at the same time highlighting the importance of \emph{entropic divergence measures} (\cref{section:entropic_divergence_measures}), as well as several other related quantum distance measures (\cref{section:quantum_distance_measures}). Next, we show in \cref{section:thermo_path_length_uncomplexity_derivation} how to derive the lower bound SWAP-gate count for the QCMP, making heavy use of quantum information geometry and circuit complexity theory \cite{yungerhalpern_uncomplexity,susskind2ndlawcomplexity,path_length1,path_length2,path_length3,measure_thermo_length,li2022wasserstein_complexity}. In \cref{section:algorithmic_implementation} we describe an algorithmic implementation for calculating the resulting SWAP uncomplexity figure of merit; we also treat several practical subtleties of the problem (\cref{section:the_beta_parameter,section:qubit_assignment}). Finally, we present the numerical results from over $600$ benchmarks tested on realistic quantum circuits and hardware layouts in \cref{section:benchmark_evaluation_and_results}; we then provide a physical interpretation of the SWAP uncomplexity using a Penrose diagram as a guide, in \cref{section:discussion}. Lastly, we offer final conclusions and ideas for future work in \cref{section:conclusion}.  

\section{Quantum Circuit Mapping} \label{section:QCMP_Background}

Generally, quantum algorithms and their associated circuit-level descriptions are developed without considering the architecture-specific limitations of particular devices, i.e., they are developed in an \emph{architecture-free} manner. For example, the \emph{elementary gate set} (or primitives) for a particular device may differ significantly from what has been indicated at the level of a generic circuit description; as such, several actions must be performed in order to translate the quantum algorithm into a circuit that a quantum device can actually execute. Another example can be seen in the physical connectivity properties of a quantum processor, which must be considered to ensure that the necessary qubit-qubit interactions of the circuit can be performed on the device. Although certain exceptions may exist (in which several of the aforementioned steps may not necessarily be carried out), these procedures are collectively known as \emph{quantum circuit mapping} \cite{bandic2020structured}.

\begin{figure*}
\centering
\includegraphics[width=0.8\textwidth]{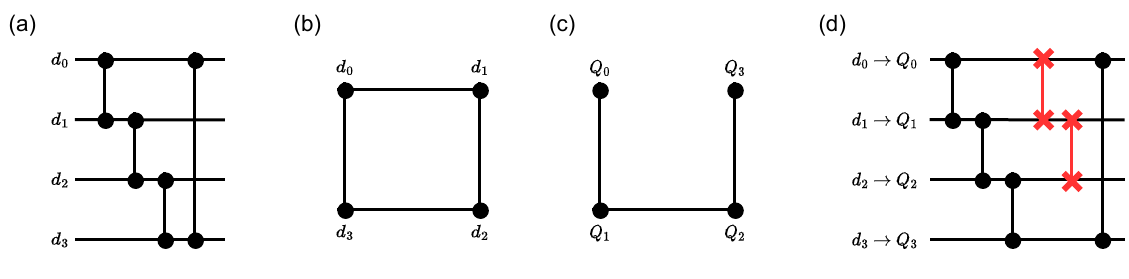}
\caption{An example of the QCMP as a sequence of steps needed to assign qubits from an algorithm to a quantum device. 
The two-qubit gates in the two circuit diagrams are used to represent general two-qubit unitary operations (with the exception of SWAP gates); here, we do not consider single-qubit gates, and the two-qubit interactions shown in (a) and (d) are taken to be general two-qubit unitary operations. (a) The quantum circuit is transformed into an \emph{interaction graph}, as shown in (b). Next, it is compared with the connectivity properties of the \emph{coupling graph} (c). As no \emph{graph isomorphism} (i.e., no exact matching between the vertices of the IG and CG which upholds all of the edge relations of both) exists between the IG and CG, one can compensate for the lack of connectivity by introducing SWAP operations to the circuit in order to realize the circuit. (d) These operations degrade the fidelity of the final output state.}
\label{figure:examplemapping}
\end{figure*}

The task of quantum circuit mapping itself is usually divided into several steps which typify the process: A) \emph{elementary gate-set decomposition}, which involves the translation of a circuit to a native gate set utilized by a quantum processor; B) \emph{scheduling}, which concerns the formation of a logical time ordering for algorithm execution, and includes considerations for parallelism of operations and for the shortening of circuit depth; C) \emph{qubit assignment}, which relates to the initial assignment of qubits from an algorithm to the physical qubits on a quantum architecture; and D) \emph{qubit routing}, which examines the increase in gate overhead as extra operations are inserted into the algorithm as a function of physically moving qubits around the processor, such that the required two-qubit operations are realizable \cite{bandic2020structured}. Typically, operations such as SWAP gates are utilized in order to adapt the circuit to hardware; these amount to classical permutation operations on a product state, but other approaches exist as well \cite{gorshkov,teleportation_wille,devulapalli2022quantum}.

As a simple example, consider the quantum circuit-mapping procedure in \cref{figure:examplemapping}. The circuit on the left is decomposed into IG form (a)-(b), wherein we do not consider single-qubit gates for simplicity, and we assume that the two-qubit gates shown in the circuit diagrams are taken as general two-qubit operations. Upon comparing the IG with the available qubit-qubit interactions afforded by a quantum device, it is apparent that the connectivity available on the device and the connectivity required by the algorithm differ (the two-qubit interaction represented by the edge $e_{d_{0}d_{1}}$ is not possible, as shown in (b)-(c)). As such, this discrepancy can be compensated for by adding detrimental SWAP operations to the circuit's initial assignment (d). This example provides an illustration of \emph{quantum circuit mapping} based on an exact graph matching between IG and CG. Implicit to this example was the assumption that exactly such a mapping between the vertices of the IG and CG exists which preserves all of the edge relationships of the IG; this is known as a \emph{graph isomorphism}, and will be discussed in more detail in \cref{section:GraphTheoryBackground}. 

In this paper, we formulate and solve a special case of the QCMP. In contrast to \cref{figure:examplemapping}, our formalism exhibits three main simplifications. Firstly, as is typical in the QCMP, we do not consider single-gate interactions in the formalism that we present starting in \cref{section:graphs_as_density_matrices}. Secondly, we assume the existence of a noiseless quantum device, as such an ideal scenario precisely allows for the emergence of a lower bound. Thirdly, we assume that no further gate simplifications via \emph{quantum circuit-synthesis techniques} can be further leveraged \cite{circuit_synthesis,synth1,synth2}. Finally, in real quantum hardware, typically we only consider that one multiqubit gate can be performed on a given hardware qubit during a given moment; this necessitates the division of the quantum circuit into time slices. Instead, we consider the scenario in which all two-qubit gate interactions of the circuit can be performed within a single unit time slice, i.e. the causal structure of our circuits are taken to be indefinite \cite{brukner2014quantum,goswami2020experimentsquantumcausality}, as essentially infinite parallelization of two-qubit gate operations can be implemented. We devote more detail to these concepts in \cref{section:discussion}.

\section{Graphs as Density Matrices}\label{section:graphs_as_density_matrices}

In quantum physics, the most general manner of describing a quantum state involves the use of \emph{density matrices} \cite{nielsenchuang,sakurai1995modern}. A density matrix $\bm{\rho}$ is a Hermitian, positive semidefinite matrix, whose trace is equal to unity. A system $\bm{\rho}$ is termed \emph{pure} if and only if the bound $\Tr[\bm{\rho}^{2}] \leq 1$ is saturated. The density matrix admits a spectral decomposition as 

\begin{equation}
\bm{\rho} = \sum_{j} p_{j} \ket{\psi_{j}}\bra{\psi_{j}}~,
\end{equation}

for an orthonormal basis $\{\ket{\psi_{j}}\}$, where $p_{j}$ are non-negative eigenvalues summing to $1$.




In this work, as in \cite{biamonte_spectral,biamonte_nature}, we make use of the concept of a density matrix to describe a complex network (i.e. a graph with many edges and vertices, and assumed topological structure \cite{strogatz2001exploring}), by defining a matrix from a network which fulfills the mathematical properties of a density matrix. One such candidate was previously shown in \cite{biamonte_spectral} to be promising for adhering to the property of \emph{subadditivity} for the VNE; this equilibrium \emph{Gibbs state} is defined as

\begin{equation} \label{eq:gibbs_state}
\bm{\rho}_{L} = \frac{e^{-\beta \mathbf{L}}}{Z}~, 
\end{equation}

where $\bm{\rho}_{L}, e^{(\cdot)}, \beta,$ and $Z$ represent: the density matrix of graph Laplacian $\mathbf{L}$; the matrix exponential; the inverse temperature (or diffusion time \cite{asymptoticentropy_diffusiontimepaper}); and the partition function, which is defined as $Z = \Tr\big[ e^{-\beta \mathbf{L}} \big]$, respectively. We define the graph Laplacian as $\mathbf{L} := \mathbf{D} - \mathbf{A}$, following \cite{graph_th_book,biamonte_spectral}. Throughout the text, we will refer to the graph Laplacians for the IG and CG using the notation $\mathbf{L}_{\text{IG}}, \mathbf{L}_{\text{CG}}$, respectively, and $\bm{\rho}_{{\text{IG}}}, \bm{\sigma}_{\text{CG}}$ to refer to the corresponding IG and CG density-matrix forms, respectively. Additionally, we refer to edges in a graph-theoretic context as a line connecting two vertices; in the density-matrix formalism, we will make reference to this instead with the term \emph{subsystem interactions}.

Using these objects to describe complex networks is advantageous for several reasons. Firstly, although it is known that the graph Laplacian is uniquely determined up to vertex-numbering assignments \cite{algebraic_godsil_graph_theory}, the \emph{eigenvalue spectrum} of the graph Laplacian does not allow for unique identification of a graph. For example, two graphs can be \emph{cospectral}, i.e. possessing the same eigenvalue spectrum, but with different connectivity \cite{algebraic_godsil_graph_theory}. As such, the approach we detail in this work is motivated by the fact that \emph{entropic divergence measures} allow for a unique differentiation between two quantum states $\bm{\rho}$ and $\bm{\sigma}$. Secondly, the VNE is \emph{permutation-invariant}, i.e., the VNE is invariant under a reordering of subsystems. For example, suppose we have a state vector of five subsystems $\mathtt{a},\mathtt{b},\mathtt{c},\mathtt{d},$ and $\mathtt{e}$. If two such subsystem orderings give rise to density matrices $\eta = \sum_{\mathtt{abcde} \in \mathbb{Z}_{2}}\ket{\mathtt{abcde}}\bra{\mathtt{abcde}}$ and $\xi = \sum_{\mathtt{abcde} \in \mathbb{Z}_{2}}\ket{\mathtt{baced}}\bra{\mathtt{baced}}$, it can be shown that the equality $\mathcal{S}(\eta) =  \mathcal{S}(\xi)$ holds \cite{nielsen_majorization_book}. Lastly, as discussed in \cite{biamonte_nature,biamonte_spectral}, previous attempts to calculate the classical entropy of a complex network fail, as these measures are dependent on a probability distribution resultant from a specific network descriptor. In contrast, the quantum approach we utilize does not depend on a specific network descriptor, but rather the entire network, rescaled and normalized as a Gibbs state.

\subsection{Distance Measures in Quantum Information Theory}\label{section:quantum_distance_measures}

The task of distinguishing two quantum states is in general a highly non-trivial problem in quantum physics, with many interpretations useful for distinct scenarios \cite{nielsenchuang,wilde_qit,watrous2018theory,li2022wasserstein_complexity}. However, at the core of these distance measures lies a central object known as the \emph{quantum Fisher information} \cite{liu2020quantumFI}, and is typically calculated as 

\begin{equation}
\mathsf{G}_{ij} = \sum_{i,j=0}^{d-1} \frac{Re \big( \bra{\lambda_{i}}\partial_{i}\bm{\rho}\ket{\lambda_{j}}\bra{\lambda_{j}}\partial_{j}\bm{\rho}\ket{\lambda_{i}}\big)}{\lambda_{i} -\lambda_{j}}~,
\end{equation}

where $\partial_{i} = \partial \rho / \partial i$ for some density matrix parameterized by a vector $\bar{\theta} = \{ \theta_{1} \dots \theta_{m}\}$, and we write $i, j$ as shorthand for the parameters $\theta_{i}, \theta_{j}$. $\lambda_{i}, \lambda_{j}$ represent the eigenvalues associated with $\partial_{i}\bm{\rho}$ and $\partial_{j}\bm{\rho}$. 

The quantum Fisher information is a fundamental object in quantum information theory, allowing for the derivation of an extended family of statistical inference measures that distinguish between parameterized quantum states in different settings \cite{jarzyna2020geometric_stat_inference,bengtsson2017geometry_of_quantum_states_book}. In particular, it is known that the quantum Fisher information is closely related to the \emph{Bures distance} $\mathcal{B}_{ij}$ \cite{nielsenchuang,watrous2018theory,bengtsson2017geometry_of_quantum_states_book}, as well as to the \emph{quantum relative entropy} as

\begin{equation}
\mathsf{G}_{ij} = 4\mathcal{B}_{ij} = 8 \big( 1 - \sqrt{\mathcal{F}(\bm{\rho_{i}},\bm{\rho_{j}})} \big) \approx \sqrt{2\mathcal{S}(\bm{\rho_{i}}|| \bm{\rho_{j}})}~,   \label{eq:equivalence_qfmi_qjsd}
\end{equation}

where $\mathcal{F}(\bm{\rho_{i}},\bm{\rho_{j}}) = \big( \text{Tr}\big[ \sqrt{\sqrt{\bm{\rho_{j}}} \bm{\rho_{i}} \sqrt{\bm{\rho_{j}}}} \big] \big)^{2} $ represents the \emph{fidelity function}, $\mathcal{S}(\bm{\rho_{i}}|| \bm{\rho_{j}}) = \mathcal{S}(\bm{\rho_{i}}) + \mathcal{S}(\bm{\rho_{j}})$ is the \emph{quantum relative entropy} (QRE), and $\mathcal{S}(\bm{\rho_{i}})$ is the \emph{Von Neumann entropy} (VNE). 

From the quantum relative entropy, one can immediately define a similar divergence measure which will be useful to the present work, the \emph{quantum Jensen-Shannon divergence}. It is defined, using the quantum relative entropy, as 

\begin{equation}
\mathcal{D}_{qJSD}(\bm{\rho_{i}}||\bm{\rho_{j}}) = \frac{1}{2} \big[ \mathcal{S}\big( \bm{\rho_{i}}|| \frac{\bm{\rho_{i}} +\bm{\rho_{j}}}{2}\big) + \mathcal{S}\big( \bm{\rho_{j}}|| \frac{\bm{\rho_{i}} +\bm{\rho_{j}}}{2} \big) \big]~.
\end{equation}

It is also well-known that the quantum Fisher information (and by extenstion, the quantum Jensen-Shannon divergence) is closely related to the \emph{quantum Wasserstein distance} \cite{li2022wasserstein_complexity,path_length1,datta2020relating}.

\subsection{Quantum Information Theory \& Entropic Divergence Measures}\label{section:entropic_divergence_measures}

The task of actually distinguishing two quantum states $\bm{\rho_{i}}$ and $\bm{\rho_{j}}$ can be accomplished through the use of the \emph{entropic divergence measures} \cite{metriccharacter_qJSD,nielsen_majorization_book,distinguishability_qJSD_paper,properties_qJSD,watrous2018theory,de2015structural,nielsenchuang,jarzyna2020geometric_stat_inference,bengtsson2017geometry_of_quantum_states_book}. In particular, we employ the \emph{quantum Jensen-Shannon divergence} (qJSD), which is defined as 

\begin{equation}
\mathcal{D}_{qJSD}(\bm{\rho_{i}}||\bm{\rho_{j}}) = \mathcal{S}\bigg( \frac{\bm{\rho_{i}}+\bm{\rho_{j}}}{2} \bigg) - \frac{1}{2}\big( \mathcal{S}(\bm{\rho_{i}}) + \mathcal{S}(\bm{\rho_{j}})\big)~.
\end{equation}

Here we defined the \emph{Von Neumann entropy} \cite{nielsen_majorization_book,nielsenchuang} as 

\begin{equation}
\mathcal{S}(\bm{\rho_{i}}) = -\Tr(\bm{\rho_{i}} \log{\bm{\rho_{i}}})~,
\end{equation}

where all logarithms are of natural base. The VNE exhibits several interesting properties: \\

\begin{itemize}[nolistsep,noitemsep]
\item \textbf{\emph{Permutation-invariance} with respect to subsystem ordering:} that is, given a multipartite quantum state, the VNE is invariant under the specific ordering of subsystems that we choose. For example, given a state vector of five subsystems $\mathtt{a},\mathtt{b},\mathtt{c},\mathtt{d},$ and $\mathtt{e}$, we have $\mathcal{S}(\bm{\rho_{i}}) =  \mathcal{S}(\bm{\rho_{j}})$, if $\bm{\rho} = \sum_{\mathtt{abcde} \in \mathbb{Z}_{2}}\ket{\mathtt{abcde}}\bra{\mathtt{abcde}}$, and $\bm{\rho_{j}} = \sum_{\mathtt{abcde} \in \mathbb{Z}_{2}}\ket{\mathtt{baced}}\bra{\mathtt{baced}}$. \\
\item \textbf{Unitary-transformation invariance:} a density matrix $\bm{\rho_{i}}$ is invariant under $\mathcal{S}(\bm{\rho_{i}}) = \mathcal{S}(U\bm{\rho_{i}} U^{\dagger})$, where $U$ is a unitary transformation. \\
\item \textbf{Additivity:} the VNE is additive for independent systems. For example, for independent subsystems $i$ and $j$, we have $\mathcal{S}(\bm{\rho_{i}} \otimes \bm{\rho_{j}}) = \mathcal{S}(\bm{\rho_{i}}) + \mathcal{S}(\bm{\rho_{j}})$. \\
\item \textbf{Subadditivity:} crucially, we see that a conjoined system $ij$ has $\mathcal{S}(\bm{\rho_{ij}}) \leq \mathcal{S}(\bm{\rho_{i}}) + \mathcal{S}(\bm{\rho_{j}})$. \\
\end{itemize}

These properties are immediately useful for defining the entropic divergence measures which were introduced in \cref{section:quantum_distance_measures}, and also are desirable for the QCMP.  

One may ask why we chose to utilize the qJSD, and not other quantum entropic measures, such as the mutual information or the quantum relative entropy \cite{nielsenchuang,jarzyna2020geometric_stat_inference}. Our deference to the qJSD is due to several useful properties (partially originating from the VNE), but arguably the most important one originates from the square root of the qJSD lies in a \emph{metric space} $\mathfrak{D}(x,y)$ for two objects $x,y$ that we wish to distinguish. A metric space is endowed with the properties of: \\

\begin{itemize}[nolistsep,noitemsep]
\item \textbf{Distance:} Let $x,y,z$ be the elements inside a set $X$, then the function $\mathfrak{D}: X \times X \mapsto \mathbb{R}$ upholds $\mathfrak{D}(x,y) \geq 0$, with the case of $\mathfrak{D} = 0$ if $x = y$. \\
\item \textbf{Symmetricity:} The function $\mathfrak{D}(x,y)$ also obeys $\mathfrak{D}(x,y) = \mathfrak{D}(y,x)$. \\
\item \textbf{Adherence to the Triangle inequality:} lastly, $\mathfrak{D}(x,y) + \mathfrak{D}(x,z) \geq \mathfrak{D}(y,z)$. \\
\end{itemize}

If these conditions are all upheld, we say that $\mathfrak{D}(\cdot,\cdot)$ is a metric space \cite{metriccharacter_qJSD}. More specifically, the qJSD defines a bounded metric space of the form

\begin{equation}
0 \leq \sqrt{\mathcal{D}_{qJSD}(\bm{\rho_{i}}||\bm{\rho_{j}})} \leq 1~,
\end{equation}

with a value of $0$ signifying that $\bm{\rho_{i}} = \bm{\rho_{j}}$, and a value of $1$ used for the case of $\bm{\rho_{i}} \perp \bm{\rho_{j}}$ \cite{distinguishability_qJSD_paper,rossi2013attributedQJSD}. As we are comparing the density matrices related to the IG and CG of a quantum circuit and processor, it is imperative to understand the closeness of one to the other, using some bounded distance measure. As a contrasting incentive, consider measuring the quantum relative entropy of two orthogonal states; in this case, the divergence is unbounded and gives $\mathcal{S}(\bm{\rho_{i}}||\bm{\rho_{j}}) \mapsto \infty$ \cite{nielsenchuang}. In the practical setting of the QCMP, such a measure is therefore not useful and does not convey the necessary distance information. 

In addition to the metric space property, the qJSD is \emph{symmetric}. This property is concomitant to the previous property related to metric spaces, but we address it here separately. Symmetricity means that the qJSD obeys the relation 

\begin{equation}
\mathcal{D}_{qJSD}(\bm{\rho_{i}}||\bm{\rho_{j}}) = \mathcal{D}_{qJSD}(\bm{\rho_{j}}||\bm{\rho_{i}})~.
\end{equation}

For the QCMP, we observe that this relation is desirable, as we wish for the notion of distance between two density matrices to stay the same, regardless of whether one is derived from $\bm{L}_{\text{IG}}$ or $\bm{L}_{\text{CG}}$. As we will see in \cref{section:thermo_path_length_uncomplexity_derivation}, it is in fact this distance quantity that we relate to the \emph{quantum circuit uncomplexity} \cite{yungerhalpern_uncomplexity,susskind2ndlawcomplexity}. Additionally, the concept of symmetricity is paramount, as it permits us to directly relate the qJSD back to the \emph{quantum Fisher information}; indeed, it was shown using the quantum Fisher information that the qJSD exactly calculates the \emph{thermodynamic path length} between two equilibrium quantum states, and lower bounds their divergence on a Riemannian manifold \cite{measure_thermo_length,path_length1,path_length2,path_length3,datta2020relating,jarzyna2020geometric_stat_inference,bengtsson2017geometry_of_quantum_states_book}.

Finally, we note that the qJSD is \emph{non-increasing} under the action of a CP map \cite{distinguishability_qJSD_paper}, which can be formally stated as

\begin{equation}
\mathcal{D}_{qJSD}(\bm{\rho_{i}}||\bm{\rho_{j}}) \leq \mathcal{D}_{qJSD}(\Lambda(\bm{\rho_{i}})||\Lambda(\bm{\rho_{j}}))~,
\end{equation}

where $\Lambda(\cdot)$ represents the superoperator of a \emph{quantum operation}. The most general form of a quantum operation can be written in several representations; in this work we will concentrate on the Kraus representation (also known as the \emph{operator-sum} representation), stated as

\begin{equation}
\Lambda(\cdot) := \sum_{i} E_{i}\cdot E_{i}^{\dagger}~,
\end{equation}

where $E_{i}$ is the $i^{\text{th}}$ term in the sum of operators, and $\Lambda(\cdot)$ is taken to be a general quantum operation superoperator, constrained to the \emph{completely-positive} (CP) condition \cite{watrous2018theory,wilde_qit,nielsenchuang}.

We also introduce here the class of \emph{doubly-stochastic} (DS) quantum channels, with the term \emph{quantum channel} distinguishing from \emph{quantum operation} in that, in addition to the CP constraint, we additionally impose \emph{trace preservation} (TP) \cite{nielsenchuang}. In this work, we will refer to CPTP maps using $\Phi(\cdot)$. Moreover, doubly-stochastic quantum channels are \emph{unital}, meaning that the fixed point of the channel upholds the equality $\Phi(\mathbb{I}_{n}) = \mathbb{I}_{n}$ \cite{wolf_qchannels,nielsen_majorization_book}. In defining the class of doubly-stochastic quantum channels, we use the fact that any Kraus operator can be factorized, as all systems of Kraus operators implementing a quantum operation are related by a unitary transformation. A particular decomposition can be defined as 

\begin{equation}
E_{i} = \sum_{j} \sqrt{\theta_{j}}P_{j}~. \label{eq:kraus_decomp}
\end{equation}

Here $P_{j} \in \mathbb{P}_{n}$ refers to permutation matrices from the set of $n \times n$ permutation matrices, and $\theta_{j}$ refers to a probability distribution \cite{nielsen_majorization_book} (we have also omitted the indices $i$ on the right-hand side for clarity). The existence of this class of convex decomposition comes from the \emph{Birkhoff-Von-Neumann Theorem} \cite{watrous2018theory,sagawa_majorization_book,nielsen_majorization_book} for which it is known that such a decomposition can be found in polynomial time \cite{johnson1960BVNalgorithm}. 

Lastly, we present \emph{projective measurements} for density matrices constructed from graph Laplacians. Following the treatment of \cite{severini}, we define a set of orthogonal projectors $\Pi_{k}$ such that $\sum_{k}\mathcal{M}_{k} = \mathbb{I}_{n}$. The post-measurement state of a general density matrix is then

\begin{equation}
\mathcal{M}(\bm{\rho}) = \frac{\mathcal{M}_{k}\bm{\rho}\mathcal{M}_{k}}{\Tr[\mathcal{M}_{k}\bm{\rho} ]}~,
\end{equation}

where $\Tr[\mathcal{M}_{k}\bm{\rho}]$ represents the probability of the $k^{\text{th}}$ measurement outcome. Note that projective measurements are known as a specific example of a CP map \cite{wilde_qit,watrous2018theory,nielsenchuang}. In \cref{section:thermo_path_length_uncomplexity_derivation}, we shall use projective measurements to erase subsystem interactions from the density-matrix form of the IG, $\bm{\rho}_{\text{IG}}$, as well as for selecting appropriate subsystem permutations of the CG density matrix $\bm{\sigma}_{\text{CG}}$. 

\section{Thermodynamic Path Length \& (Un)Complexity}\label{section:thermo_path_length_uncomplexity_derivation}

As mentioned in \cref{section:quantum_distance_measures}, the quantum Fisher information defines an entire family of statistical distance measures, from which we have taken particular interest in the family of entropic divergences. However, it is still not clear how to connect this to the more profound notion of \emph{thermodynamic path length}. In order to provide an answer, let us start from quantum thermodynamics \cite{path_length1,path_length2,path_length3}: the distance between any two quantum states (in continuum spacetime) can always be described as the average work extracted over some path through configuration space $\xi$: 

\begin{equation}
W = \int_{\xi} dt \text{Tr}\big[ \dot{H_{t}}\bm{\rho}_{t} \big]~,
\end{equation}

where we consider $H_{t} = \sum_{i} \lambda_{i}X_{i}$ as a time-dependent Hamiltonian with time-dependent, experimentally-controllable parameters $\lambda_{i}$ and time-independent observables $X_{i}$, and $\bm{\rho}_{t}$ is the time-evolved density matrix from $t \in [0, \tau]$, following the work of \cite{path_length1,path_length2,path_length3}. 

We also generally know how $\bm{\rho}_{t}$ evolves in spacetime via the \emph{Lindblad master equation} \cite{path_length2,path_length3}, which is given by 

\begin{equation}
\frac{d\bm{\rho}}{dt} = i/\hbar \big[ H, \bm{\rho} (t) \big] + \mathcal{L}(\bm{\rho} (t))~, \label{eq:startpoint_uncomplexity}
\end{equation}

where $\mathcal{L}(\bm{\rho} (t)) = \sum_{j} \gamma_{j} \big( L_{j}\bm{\rho} L_{j}^{\dagger} - 1/2 \{ L_{j}^{\dagger}L_{j}, \bm{\rho}\} \big)$, and $\bm{\rho}_{t}, \bm{\rho} (t)$ represent the time-evolved density matrix at time $t$, as well as the time-dependence of the density matrix on $t$, respectively. $L_{j}$ are known as \emph{jump operators}, and describe the channel that the quantum system is subjected to as it interacts with external environmental degrees of freedom \cite{path_length3}. $\gamma_{j}$ are known as the \emph{decoherence rates}. 

In the case of the QCMP, this machinery is not needed, as we simply wish to understand the optimal case of quantum circuit mapping. That is to say, we wish to consider a noiseless quantum processor, capable of infinite parallelization (as we described above and in the manuscript's discussion section). In that case, there are no environmental factors nor decoherence to consider, and we can examine our problem from the standpoint of a closed quantum evolution; therefore, we set $\gamma_{j} = 0$ and recover the original Von Neumann equation. As is expected for a closed system of pure quantum states, the dynamics now depend only on the Hamiltonian.

In order to calculate thermodynamic path length, \cite{path_length3} considers the amount of work dissipated into the environment due to restricted thermodynamic transformations on Gibbs states. One can directly find this from \cref{eq:startpoint_uncomplexity}, by optimizing the geodesic equations and accompanying Christoffel symbols for $\lambda_{i}$ \cite{path_length3}, ending with 

\begin{equation}
W_{\text{diss}} = 1/\beta \int_{\xi} dt \dot{\lambda_{i}} (\mathsf{G}_{ij})_{t} \dot{\lambda_{j}}~, \label{eq:integral_path_length1}
\end{equation}

Where $\beta$ here represents the inverse temperature related to the Gibbs state, as is standard. From $W_{\text{diss}}$, we can formally define thermodynamic path length as

\begin{equation}
\mathsf{W}_{\text{path}} = 1/\beta \int_{\xi} dt \sqrt{\dot{\lambda_{i}} (\mathsf{G}_{ij})_{t} \dot{\lambda_{j}}}~. \label{eq:integral_path_length2}
\end{equation}

In these previous two equations, we recognize the quantum Fisher information $\mathsf{G}_{ij}$ in the integral kernel. As we discussed previously in \cref{section:quantum_distance_measures}, it is known that thermodynamic path length and the Jensen-Shannon divergence compute the lower-bound distance between quantum states, and that this distance constitutes a geodesic in a configuration space of allowed transformations between equilibrium states \cite{liu2020quantumFI,path_length1,path_length2,path_length3,datta2020relating,li2022wasserstein_complexity,measure_thermo_length}. Lastly, we know that geodesics not only represent the shortest paths on a Riemannian manifold, but also that the distance between any two infinitesimally small intervals on the geodesic are locally the shortest path as well (i.e. the distance function we have defined must monotonically decrease along the thermodynamic path).

Before progressing, there are a few further points to mention. Firstly, when two CG vertices are adjacent to one another such that a corresponding IG two-qubit edge (gate) may be performed, we say that this edge is executable, and therefore should not factor further into the shortest-path calculation. Therefore, we compensate for this by performing \emph{Bell measurements} on these edges, an operation already shown in \cite{severini} to erase edges of simple graphs. We use the measurement scheme in quantum-operation form, as discussed at the end of \cref{section:entropic_divergence_measures}. Secondly, in order to permute vertices on the CG, we make extensive utilization of the \emph{doubly-stochastic} quantum channel forms which are also described in \cref{section:entropic_divergence_measures}. However, we must make a slight modification due to the practical considerations of the QCMP. As we are limited to performing only nearest-neighbor SWAP gates on the CG, this signifies that, for the Kraus operators $E_{i} = \sum_{j} \sqrt{\theta_{j}} P_{j}$, we have that $P_{j} \in \mathbb{P}_{n}(\text{CG}) \subset \mathbb{P}_{n}$, where $\mathbb{P}_{n}(\text{CG})$ is the subgroup of all nearest-neighbor permutations available on the CG at a given time instant. 

Thirdly and lastly, the operations of a quantum circuit take place over the discrete configuration space of $SU(2^{k})$, with $k=2,3$ in most contexts, representing the number of qubits participating in a given gate \cite{susskind2ndlawcomplexity}. Although it is the case that $SU(2^{k})$ is a Lie group and is therefore continuous, the permitted operations of the QCMP lie strictly within discrete configuration space, as we are restricted to only the set of Bell measurements $\{\mathcal{M}_{e}\}_{e \in E_{\text{IG}}}$ on the IG, and doubly-stochastic quantum operations $\{ \Lambda(\cdot) | E_{i} \in \mathbb{P}_{n}(\text{CG}) \}$ on the CG, both of which are represented by discrete simple graphs. 

We can then discretize the integral over $\xi \in  \langle \{\mathcal{M}_{e}\}_{e \in E_{\text{IG}}}, \Lambda(\cdot) \rangle$ by considering infinitesimally small time translations $t + \Delta t$ with $\Delta << 1$, such that 

\begin{equation}
\bm{\rho} (t + \Delta t) \approx \mathcal{E}(\bm{\rho} (t))~, 
\end{equation}

where $\mathcal{E}(\cdot)$ is the action of a quantum channel acting on $\bm{\rho}$. If we then perform this action $m$ times, then we have \\

\begin{equation}
\bm{\rho} (t + m(\Delta t)) \approx \mathcal{E}_{m} \circ \cdots \circ \mathcal{E}_{1}(\bm{\rho} (t))~.
\end{equation}

Keeping all of these points in mind, we can re-write the integral from \cref{eq:integral_path_length1} in discretized form as 

\begin{equation}
W_{\text{diss}} = \sum_{m} \mathsf{\Pi}^{l} \big[ \mathcal{M}^{l}_{i} \big[ \argmin_{\mathsf{G}_{ij}} \big[ \mathsf{\Pi}^{m}\big[ \Lambda^{m}_{j}(\mathsf{G}_{ij})\big] \big] \big] \big]~,
\end{equation}

where $\mathsf{\Pi}^{l}\big[ \mathcal{M}^{l}_{i} (\cdot) \big] = \mathcal{M}^{l}_{i} \circ \cdots \circ \mathcal{M}^{1}_{i} (\cdot)$, i.e. the sequence of measurements executed on the IG when two-qubit gates are possible on the CG. Additionally, $\mathsf{\Pi}^{m}\big( \Lambda^{m}_{j} \big( \cdot \big)\big) = \Lambda^{m}_{j} \circ \cdots \circ \Lambda^{m}_{j} \big( \cdot \big)$, i.e. the sequence of SWAP-gate permutations undertaken in order to move qubits on the CG such that IG two-qubit gates can be performed. We sum over all of the permutations performed as we are erasing IG edges. Finally, as we recognize from \cite{measure_thermo_length,datta2020relating,li2022wasserstein_complexity} and \cref{eq:equivalence_qfmi_qjsd} that the quantum Jensen-Shannon divergence is directed related to the quantum Fisher information matrix, we can directly substitute and obtain the form 

\begin{equation}
W_{\text{diss}} = \sum_{m} \mathsf{\Pi}^{l} \big[ \mathcal{M}^{l}_{i} \big[ \argmin_{\mathcal{D}^{\text{qJS}}_{ij}} \big[ \mathsf{\Pi}^{m}\big[ \Lambda^{m}_{j}(\mathcal{D}^{\text{qJS}}_{ij}(\bm{\rho}_{\text{IG}} || \bm{\rho}_{\text{CG}}))\big] \big] \big] \big]~,
\end{equation}

where $\mathcal{D}^{\text{qJS}}_{ij}(\bm{\rho}_{\text{IG}} || \bm{\rho}_{\text{CG}}) = \mathcal{S}(\frac{\bm{\rho}_{\text{IG}} + \bm{\rho}_{\text{CG}}}{2}) - \frac{1}{2} \big( \mathcal{S}(\bm{\rho}_{\text{IG}}) + \mathcal{S}(\bm{\rho}_{\text{CG}}) \big)$, and the subscripts $i,j$ denote quantum operations on the IG and CG, respectively. We have also absorbed the terms $\dot{\lambda_{i}}$ and $\dot{\lambda_{j}}$ into the description of their respective quantum channels, as these terms represent time-dependent, externally-controllable parameters in the first place. The equation above can be likened to the process of \emph{parallel transport} on a Riemannian manifold \cite{datta2020relating,li2022wasserstein_complexity}, and preserves the structure of the metric, as well as the geodesic form. \\ 

As we sum over all of the $m$ permutations performed, we eventually erase all of the edges of the IG, resulting in an effective distance between the original CG and the maximally-mixed state (now the erased IG):

\begin{multline}
W_{\text{diss}} = || \mathbb{I}_{\text{IG}} - \mathsf{\Pi}^{m}\big[ \Lambda^{m}_{j}(\bm{\rho}_{\text{CG}})\big] ||_{\mathcal{O} \in \langle \{\mathcal{M}_{e}\}_{e \in E_{\text{IG}}}, \Lambda(\cdot) \rangle} \\ = \mathcal{C}(\mathbb{I}_{\text{IG}}) - \mathcal{C}(\rho_{\text{CG}}) = \mathsf{U}_{\text{SWAP}}~, \label{eq:uncomplexity_swap}
\end{multline}

where $|| \cdot ||_{\mathcal{O}}$ is a distance measure subject to the restrictions on transformations $\mathcal{O}$ for transporting $\mathcal{D}^{\text{qJS}}_{ij}(\bm{\rho}_{\text{IG}} || \bm{\rho}_{\text{CG}})$ along the geodesic. It is obvious from the lower half equalities of \cref{eq:uncomplexity_swap} that our equation exactly coincides with the form of \emph{quantum circuit uncomplexity} given by \cite{susskind2ndlawcomplexity} and expounded upon in \cite{yungerhalpern_uncomplexity}. Additionally, \cref{eq:uncomplexity_swap} is directly related to the \emph{quantum Wasserstein distance}, another known distance measure for calculating the shortest path between two quantum states in terms of number of gates within some restricted set of allowed transformations \cite{li2022wasserstein_complexity,datta2020relating}.

\section{Algorithmic Implementation}\label{section:algorithmic_implementation}

The pseudocode for calculating the SWAP uncomplexity is shown in \cref{alg:uncomplex_pseudo}. The algorithm proceeds similarly to the mathematical derivation detailed in \cref{section:thermo_path_length_uncomplexity_derivation}. Firstly, the density matrices $\bm{\rho}_{\text{IG}}$ and $\bm{\rho}_{\text{CG}}$ are provided as inputs, and $\mathsf{U}_{\text{SWAP}}$ is set to zero. Next, we immediately calculate the qJSD in order to check if an isomorphism exists between $\bm{\rho}_{\text{IG}}$ and $\bm{\rho}_{\text{CG}}$. If none exists, then we first remove all of the edges of the IG which directly match up with edges of the CG, obtaining the density matrix $\bm{\bar{\rho}_{\text{IG}}}$. We then set a SWAP-gate counter $m$ to zero. Afterwards, a \texttt{For} loop begins with the eventual goal to erase all of the IG's edges; the process by which this happens begins with the calculation of a first qJSD $\mathtt{qjsd}_{1}$ after the $i^{\text{th}}$ and $m^{\text{th}}$ actions of measurement and doubly-stochastic quantum channels on their respective density matrices. Additionally, we calculate a second qJSD with an extra permutation applied to the CG. The optimal choice of this particular permutation requires a worst-case search over all edges of CG for each iteration. In practice, decomposing a doubly-stochastic quantum channel will result in the superposition of several possible permutation matrices \cite{nielsen_majorization_book,sagawa_majorization_book}; in order to make a hard decision, we choose to apply the permutation matrix with the maximal $\theta_{j}$ value, as shown in \cref{eq:kraus_decomp}. The reason for choosing the maximal $\theta_{j}$ lies in the fact that performing the most-likely permutation matrix at every iteration step of the algorithm allows us to follow and stay on the geodesic at every time step \cite{path_length3,measure_thermo_length}. After applying the permutation matrix, we compute the second qJSD $\mathtt{qjsd}_{2}$; if it is found that $\mathtt{qjsd}_{2} < \mathtt{qjsd}_{1}$, then we simply add one to the SWAP counter and the same process of comparing subsequent qJSDs continues until $\mathtt{qjsd}_{2} \geq \mathtt{qjsd}_{1}$. Upon arriving here, we first check to see if the current iteration of $\mathcal{M}^{i}(\bm{\bar{\rho}}_{\text{IG}})$ is equivalent to the identity matrix $\mathbb{I}_{\text{IG}}$; in this case, the algorithm is complete and we break out of the \texttt{For} loop, returning the number $m$ associated to the SWAP uncomplexity $\mathsf{U}_{\text{SWAP}}$. If $\mathcal{M}^{i}(\bm{\bar{\rho}}_{\text{IG}}) \not \equiv \mathbb{I}_{\text{IG}}$, then we continue by performing the next subsequent measurement, $\mathcal{M}^{i+1}(\bm{\bar{\rho}}_{\text{IG}})$, associated with whichever edges are currently matched up between the IG and the CG. 

At this point, the erasure of a remaining subsystem interaction implies that the qJSD will again increase, as we know that the VNE under a CP map always increases \cite{benatti1988entropy_under_CP_maps}. We must then perform the commensurate doubly-stochastic quantum channel operation(s) again and select the appropriate $\theta_{j}$-valued permutation such that the divergence decreases to its minimal value once more. The algorithm terminates upon the successful erasure of all subsystem interactions in $\bm{\bar{\rho}}_{\text{IG}}$, leaving a maximally mixed state. 

One may ordinarily surmise that the runtime complexity of \cref{alg:uncomplex_pseudo} is quite high; after all, inside the \texttt{For} loop lies several seemingly difficult optimization problems. However, due to the \emph{Birkhoff-Von-Neumann algorithm}, decomposition of any doubly-stochastic quantum channel is guaranteed in polynomial timesteps \cite{johnson1960BVNalgorithm,nielsen_majorization_book}.
Taking stock, we conclude that the algorithm's runtime complexity is bounded by the number of edges in the IG, multiplied by the number of edges in the CG queried by an optimizer to determine the maximum-$\theta_{j}$ permutation for estimating \verb|qjsd|$_2$, using $\Lambda^{m+1}$. Since $|\bar{E}_{\text{IG}}| \leq |\bar{E}_{\text{CG}}|$, the worst case complexity is quadratic in the number of edges in a complete graph of size CG, or more succinctly, $O((\text{dim}(\bm{\rho}_{\text{CG}})(\text{dim}(\bm{\rho}_{\text{CG}})-1)/2)^2) = O(\text{dim}(\bm{\rho}_{\text{CG}})^4)$. However, for the optimization loop, since we need to consider edges within the connectivity constraints of quantum processors, these graphs are typically planar instead of all-to-all connected (i.e. as in a complete graph), with much more benign runtime expectation for the pragmatic use case. 

\begin{algorithm}
\caption{Pseudocode for an algorithmic optimization of the SWAP Uncomplexity $\mathsf{U}_{\text{SWAP}}$.}\label{alg:uncomplex_pseudo}

\KwIn{$\bm{\rho}_{\text{IG}},\bm{\rho}_{\text{CG}}$}

$\text{Initial\_Qubit\_Assignment} \gets \text{Qubit\_Assignment}(\bm{\rho}_{\text{IG}},\bm{\rho}_{\text{CG}})$ \\
$\mathsf{U}_{\text{SWAP}} \gets 0$ \\

\KwOut{$\mathsf{U}_{\text{SWAP}}$}

\Assert{$\mathcal{D}_{\text{qJS}}(\bm{\rho}_{\text{IG}}||\bm{\rho}_{\text{CG}}) == 0$} \\

\If{$\mathcal{D}_{\text{qJS}}(\bm{\rho}_{\text{IG}}||\bm{\rho}_{\text{CG}}) ~!= 0$}{

$\bm{\bar{\rho}}_{\text{IG}} \gets$ Remove\_Trivial\_Edges$(\bm{\rho}_{\text{IG}})$ \\

$m \gets 0$ \\ 

\For{$i \in [0,|\bar{E}_{\text{IG}}|]$}{

$\mathtt{qjsd}_{1} \gets \mathcal{D}_{\text{qJS}}\big( \mathcal{M}^{i}(\bm{\bar{\rho}}_{\text{IG}})|| \Lambda^{m}(\bm{\rho}_{\text{CG}}) \big)$ \\

$\mathtt{qjsd}_{2} \gets \mathcal{D}_{\text{qJS}}\big( \mathcal{M}^{i}(\bm{\bar{\rho}}_{\text{IG}})|| \Lambda^{m+1}(\bm{\rho}_{\text{CG}}) \big)$ \\

\eIf{$\mathtt{qjsd}_{2} < \mathtt{qjsd}_{1}$}{

$m \gets m + 1$ \\ 
\textbf{continue} \\ 
}{
\eIf{$\mathcal{M}^{i}(\bm{\bar{\rho}}_{\text{IG}}) \equiv \mathbb{I}_{\text{IG}}$}{

\textbf{break}
}{

$\mathcal{M}^{i}(\bm{\bar{\rho}}_{\text{IG}}) \gets \mathcal{M}^{i+1}(\bm{\bar{\rho}}_{\text{IG}})$

}
}

}

}

\Return{$\mathsf{U}_{\text{SWAP}} \gets m = \mathsf{U}_{\text{SWAP}}$}{}
\end{algorithm}

\subsection{The \texorpdfstring{$\beta$}{TEXT} Parameter}\label{section:the_beta_parameter}

\begin{figure}
\centering
\includegraphics[width=\columnwidth]{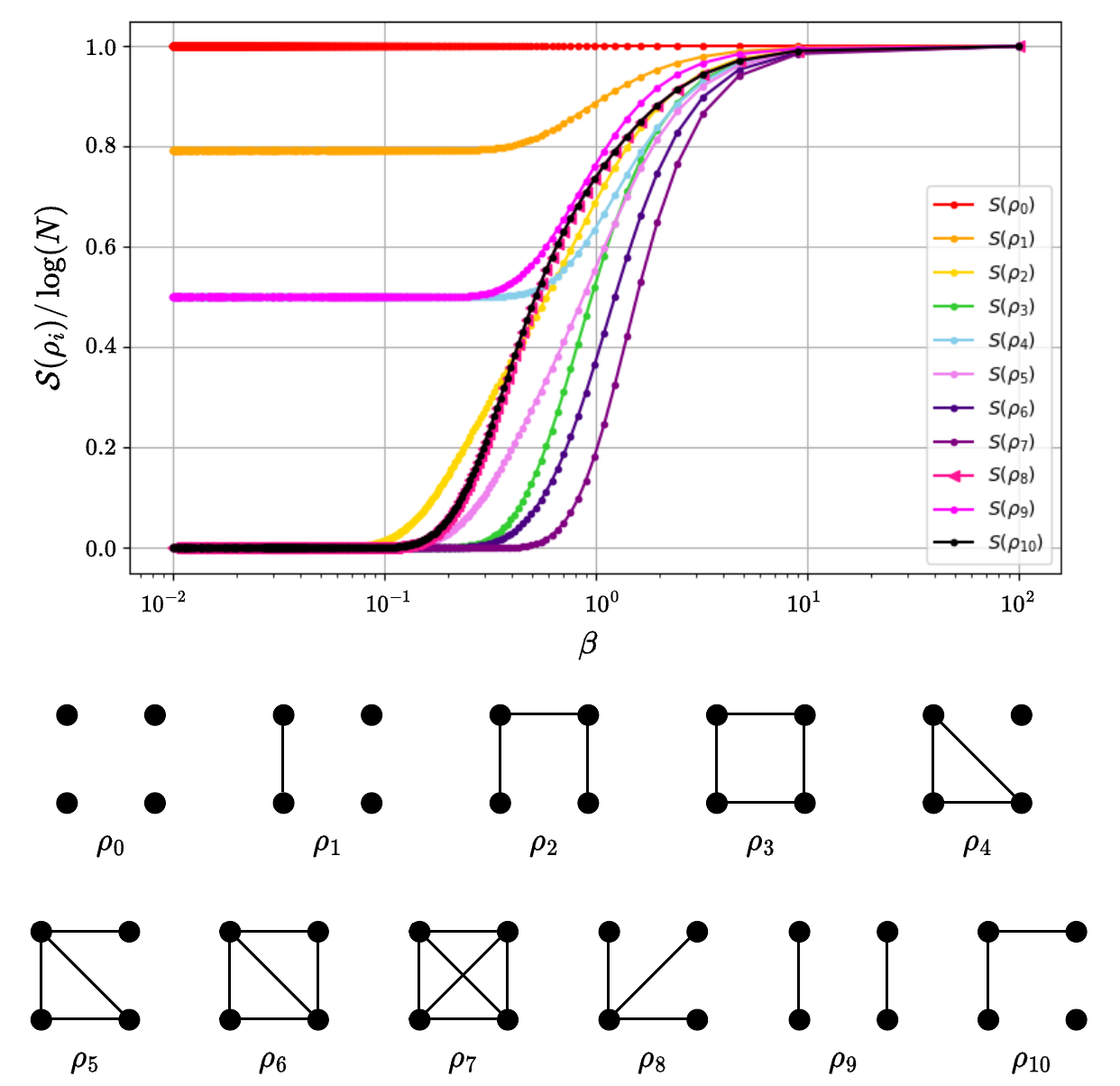}
\caption{The fluctuation of $\mathcal{S}_{G}(\bm{\rho_{i}})$ as a function of $\beta$ for $4$-node graphs. We have utilized different marker types in order to distinguish curves with very similar VNE.}
\label{figure:betaVSvne}
\end{figure}

Estimating the inverse temperature value $\beta$ in \cref{alg:uncomplex_pseudo} can be a delicate procedure. During our investigation, we uncovered that a non-trivial dependence of the VNE exists on $\beta$; this is evident in \cref{figure:betaVSvne}, where we have graphed $\beta$ versus the normalized VNE value. It is well-known in the theory of \emph{phase transitions} that rapid changes in entropy which are dependent on temperature are prime signals of phase transitions \cite{thermal_phys_book}, and this is clear in \cref{figure:betaVSvne} about the point $\beta \sim 10^{0}$. Conversely, the \emph{high-temperature regime} generally begins at $\beta \sim 10^{-1}$ and extends leftwards. A search was conducted over a range of $\beta$ values with varying increments; the preferred value yields the minimal SWAP-gate count for an IG and CG pair. We were able to ascertain this minimal SWAP-gate count using a search over the range of $10^{-5} \leq \beta \leq 10^{5}$. As the definition of SWAP uncomplexity states that \emph{there exists a minimal bound} for the number of SWAP gates at \emph{some value} of $\beta$, but does not say directly how one may find the most appropriate $\beta$, we conclude that this is consistent with our definition of $\mathsf{U}_{\text{SWAP}}$. As we show in \cref{section:benchmark_evaluation_and_results}, it so happens that almost all of our results ($\sim 97\%$) achieve the minimal SWAP-gate count in the \emph{high-temperature regime}, largely following the empirically-derived results of \cite{PRE_beta_article}. We will discuss possible improvements to this technique in \cref{section:discussion,section:conclusion}. 

\subsection{Qubit Assignment} \label{section:qubit_assignment}

As mentioned previously in \cref{section:QCMP_Background}, the initial stage of the QCMP is known as \emph{qubit assignment} (also known as initial placement, qubit allocation, or initial mapping) \cite{siraichi2018qubit,siraichi2019qubit}. This procedure plays a pivotal role in quantum circuit execution \cite{venturelli_tackling_qubit_mapping}. In our proposal for calculating the SWAP uncomplexity in the QCMP, we also must assign qubits from the IG to the CG initially in an optimal way, as this influences how many SWAP gates will be utilized. 

In \cite{siraichi2018qubit}, the concept of qubit assignment was introduced as a search for a subgraph isomorphism for an IG/CG pair. To our knowledge, this technique has not yet seen widespread implementation in practical qubit-assignment techniques, despite its potential. Instead, most existing approaches focus on alternative factors such as sequential gate flow in the circuit or the number of interactions between qubits as in \cite{bahreini2015minlp,li2019tackling,tannu2019not}. Nevertheless, some work has explored the subgraph-isomorphism concept for the QCMP \cite{siraichi2019qubit,li2020qubit,jiang2021quantum,peham2023optimal}.

Building upon the foundation of the well-known VF2 algorithm \cite{li2020qubit,cordella2004sub}, our approach to qubit-assignment searches for an exact location on the quantum device where our circuit can run without requiring additional gates. If a solution is feasible, we are left with an optimal assignment. In cases where a solution is not possible, we conduct a graph similarity search. This process involves the GED calculation and comparison of the IG to all distinct subgraphs of the same size within the CG, which opens up alternative assignment possibilities. In this fashion, we condense the search space for alternative solutions, while also highlighting the potential utility of our approach for \emph{multi-programming} applications (i.e. executing multiple circuit in parallel on a quantum device) \cite{niu2022multi}.

Let $|V_{\text{IG}}|$ be the number of qubits in the IG, and $|V_{\text{CG}}|$ be the number of physical qubits on the IG and CG, respectively. Our qubit-assignment process consists of the following steps:

\begin{enumerate}[noitemsep]
\item \emph{Preprocessing:}
\begin{enumerate}[nolistsep,noitemsep]
\item \label{step:selection} Select a quantum algorithm described as a quantum circuit and extract its IG $G_{\text{IG}}(E_{\text{IG}},V_{\text{IG}})$, where $|E_{\text{IG}}|$ represents the number of edges in the IG. 
\item Choose a quantum device to execute the circuit on and extract its CG, represented as graph $G_{\text{CG}}(E_{\text{CG}},V_{\text{CG}})$, where $|E_{\text{CG}}|$ stands for the number of edges in the CG.
\item \label{step:identify} In order to increase the efficiency of steps later on, and reduce the search space, we find all distinct subgraphs of size $|V_{\text{IG}}|$ within graph $G_{\text{CG}}$.
\end{enumerate}
\item \emph{Subgraph isomorphism using VF2 and subgraph similarity search:}
\begin{enumerate}
\item Use the VF2 algorithm to check if a subgraph isomorphism exists between graphs $G_{\text{IG}}$ and $G_{\text{CG}}$.
\begin{enumerate}
\item If a subgraph isomorphism is found, we immediately determine the location within the CG for qubit assignment.
\item \label{step:GED} When a subgraph isomorphism does not exist, we utilize the \emph{graph-edit distance} (GED) (\cref{section:GraphTheoryBackground}) to identify structurally most similar subgraph of the CG when compared to the IG. During this process, we compare IGs only to distinct subgraphs of a CG derived from \cref{step:identify}.
\end{enumerate}

\item Assign the IG to the CG in accordance with the result from the previous step.
\end{enumerate}

\item \emph{Calculating the maximal SWAP-gate count} as it depends on the qubit assignment. We describe the computation of this bound in more detail in \cref{section:maxboundcalc}.
\end{enumerate}

\begin{figure}
\centering
\includegraphics[width=\columnwidth]{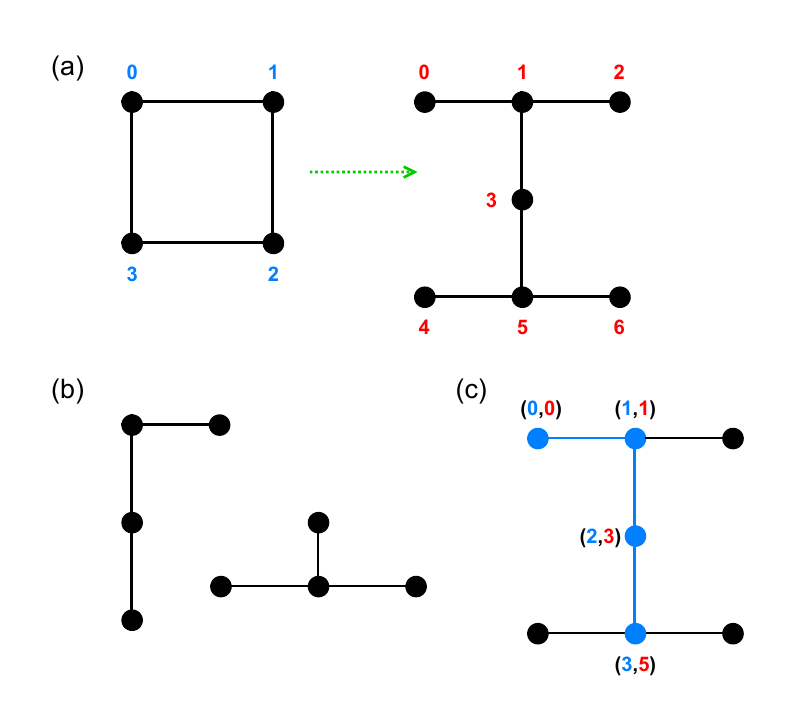}
\caption{Steps taken for the qubit-assignment algorithm described in \cref{section:qubit_assignment}: (a) The task at hand is to find the best-fit initial placement for the qubits in the $4$-qubit IG (shown with blue numbering) for the $7$-qubit CG (shown with red numbering); in our case, the the CG corresponds to the connectivity of the \emph{IBM Casablanca} quantum device. (b) Shows the distinct subgraphs found from \cref{step:identify} in \cref{section:qubit_assignment}. After verifying that no direct subgraph isomorphism between IG and one of these graphs exists, a similarity search is employed. (c) The subgraph of the CG with the lowest GED relative to an IG is retrieved. The resulting initial placement is shown in blue, with the final GED calculated to be $1$. The actual qubit assignment is shown in the form of several colored ordered pairs (blue numbering represents IG qubits, while red numbering represents CG qubits).}
\label{fig:qubassign}
\end{figure}

The steps in the algorithm are exemplified in \cref{fig:qubassign}. Here, we take as a simple example the case of a $4$-qubit IG assigned to a $7$-qubit architecture, as shown in (a). In (b), we display the two distinct $4$-qubit subgraphs that are identified in \cref{step:identify}. Finally, in (c) we show the final assignment of qubits as ordered pairs; in this case, an graph isomorphism was not found. Therefore, we select the subgraph with the lowest GED calculated, as per \cref{step:GED}. The final initial placement and related information obtained during this process serve as inputs for the SWAP uncomplexity algorithm, as described in \cref{section:thermo_path_length_uncomplexity_derivation,section:algorithmic_implementation}.

In addition to implementing the technique described above, a specific approach for \emph{complete graphs} (i.e., all-to-all IGs) was utilized. For such cases, we automatically locate the most-connected subgraph of that size within the CG. This method can also be applied to circuits larger than 20 qubits. However, for the purpose of this paper, we focus on smaller circuits as a demonstration of the concept.

\section{Benchmark Evaluation \& Results}\label{section:benchmark_evaluation_and_results}

\begin{figure*}
\centering
\includegraphics[width=0.9\textwidth]{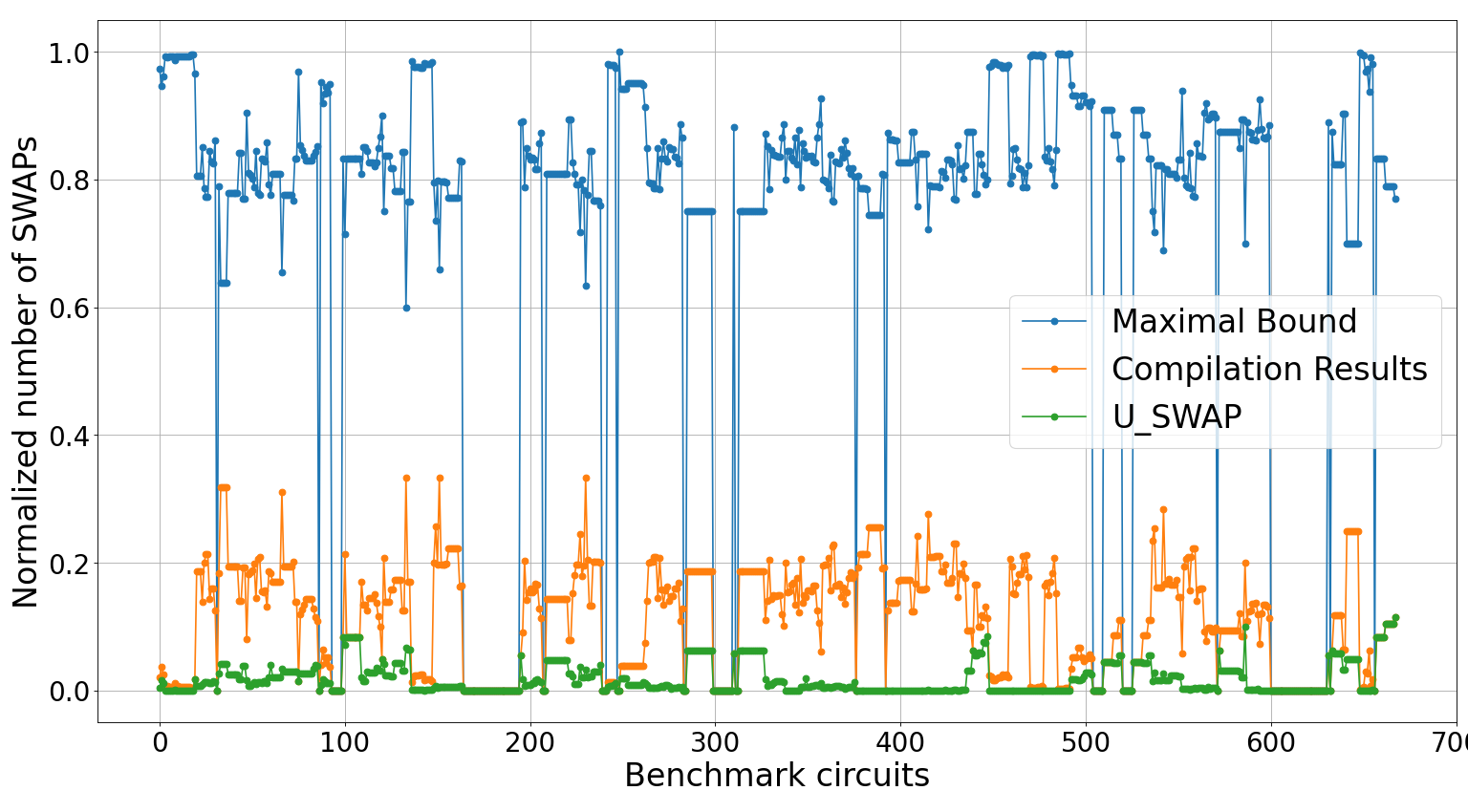}
\caption{Simulation results for various IG / CG benchmark pairs. The horizontal axis enumerates each benchmark circuit tested (sorted by the number of two-qubit gates), and the vertical axis describes the normalized number of SWAPs, due to very high maximal bounds (the SWAP-bound values of each benchmark are divided by their sum). The results are color-coded as follows: the SWAP uncomplexity of \cref{section:thermo_path_length_uncomplexity_derivation} (green); the Qiskit compiler with default options Sabre router and circuit optimization level $1$ \cite{Qiskit} (orange); and the maximal maximal bound calculated as in \cref{section:maxboundcalc} (blue). In every IG/CG pair, the bound calculated captures the SWAP uncomplexity that is either approachable or unattainable by the Qiskit compiler, thus empirically demonstrating our formulation.}
\label{figure:swapbounds}
\end{figure*}

\begin{figure}
	\centering
	\subfigure[]{\includegraphics[width=.9\linewidth]{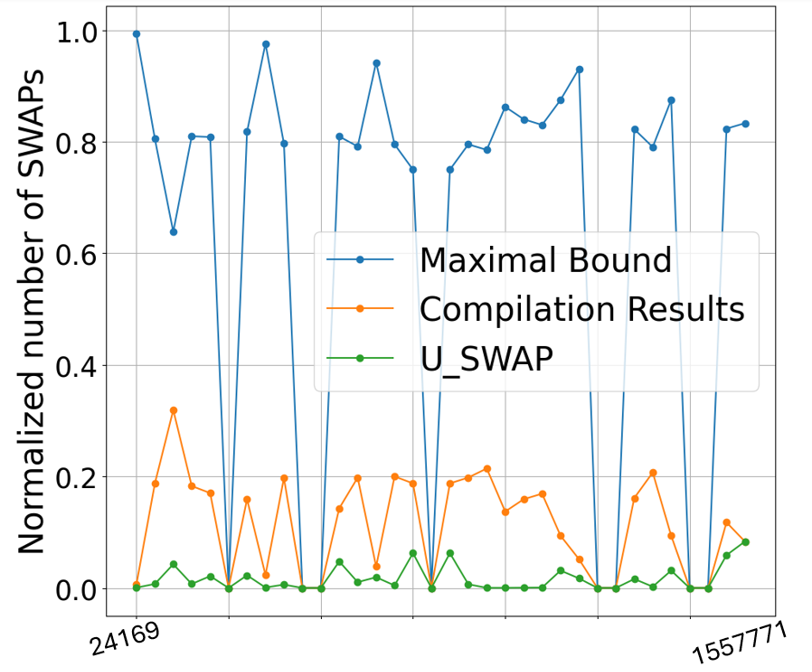}
    \label{fig:bristlecone_2qgates}} \\
    \subfigure[]{\includegraphics[width=.9\linewidth]{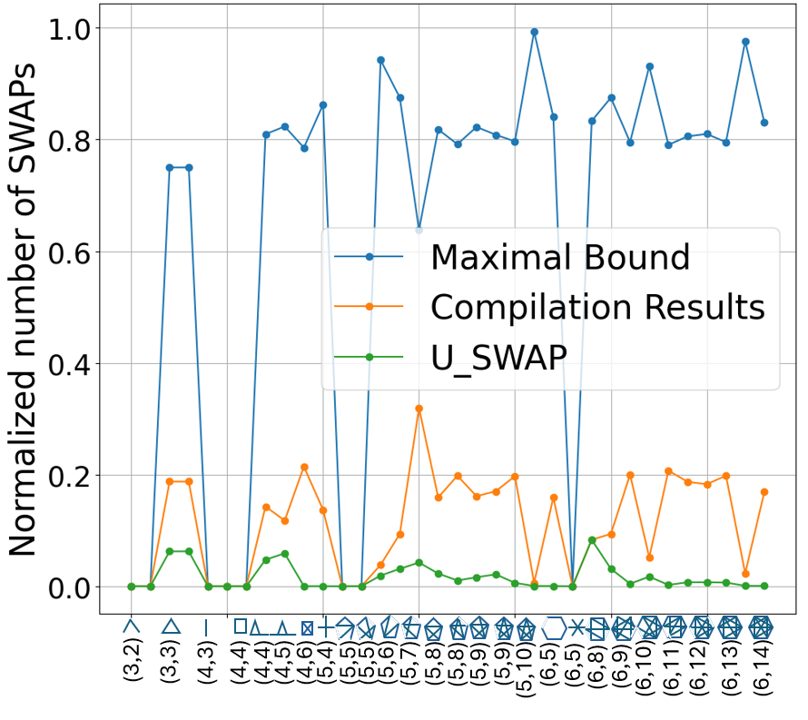}
    \label{fig:bristlecone_IG}}
    \caption{Subset of simulation results from \cref{figure:swapbounds} where only one CG is shown (Google Bristlecone). Benchmark circuits are sorted by a) number of two-qubit gates; and b) IG complexity (number of nodes and edges of IGs). The respective benchmarks with their respective nodes-edges count are detailed in \cref{tab:benchmarks}.}
\label{fig:swapbounds_bristlecone}
\end{figure}

In this section, we describe the numerical results obtained from comparing the SWAP uncomplexity against IBM's Qiskit compiler \cite{Qiskit}*, as well as against a brute-force approach \cite{BF6qubits_wille}. These experiments were carried out for two main reasons. Firstly, we wish to subject the SWAP uncomplexity formalism and algorithm to a concrete, rigorous sanity check; after all, if the SWAP uncomplexity algorithm does in fact solve for the minimal SWAP-gate count, then the bounds we calculate should not be surpassed by any known compilation or brute-force optimization method in existence. By such logic, a compiler should be able to attain but not find fewer SWAP gates for an arbitrary IG / CG pair. In order to perform this empirical check, we chose to run our algorithm (\cref{section:algorithmic_implementation}) against the Qiskit compiler, since it is considered to be the state-of-the-art approach at the moment. Secondly, to the best of our knowledge, there is scant literature on bounding required SWAP gates for an IG / CG pairing; at the moment, the latest work we are aware of addresses only up to quantum circuits of $6$ qubits via a brute-force optimization algorithm \cite{BF6qubits_wille,qubitplacement_for_wille_6qubits}. In contrast, our simulation results demonstrate scalability that greatly exceeds this brute-force optimization technique \cite{BF6qubits_wille}, as we achieved results for circuits of up to $16$ qubits.

\def\thefootnote{*}\footnotetext{Qiskit 0.24.1 was utilized in this work.}\def\thefootnote{\arabic{footnote}}

$57$ benchmark circuits were selected from the \emph{qbench} suite \cite{bandic2023interaction}. These benchmarks cover a range of $3$ to $20$ qubits and represent a wide spectrum of possible IG connectivities ($47$ different connectivities) encountered in quantum algorithms. More details about the selected benchmarks can be found in \cref{tab:benchmarks}. As for the CGs, we chose connectivity graphs from a set of $16$ in-use quantum devices, ranging from $5$ to $72$ qubits. The specific details of these devices are provided in \cref{tab:devices}. As some of the benchmarks are too large to be run on some of the smaller processors from our list, in total we devised $675$ simulation experiments with the Qiskit compiler \cite{Qiskit}. In these simulations, we utilized Qiskit's transpiler with the default circuit-optimization setting. 

\begin{figure}
\centering
\includegraphics[width=\columnwidth]{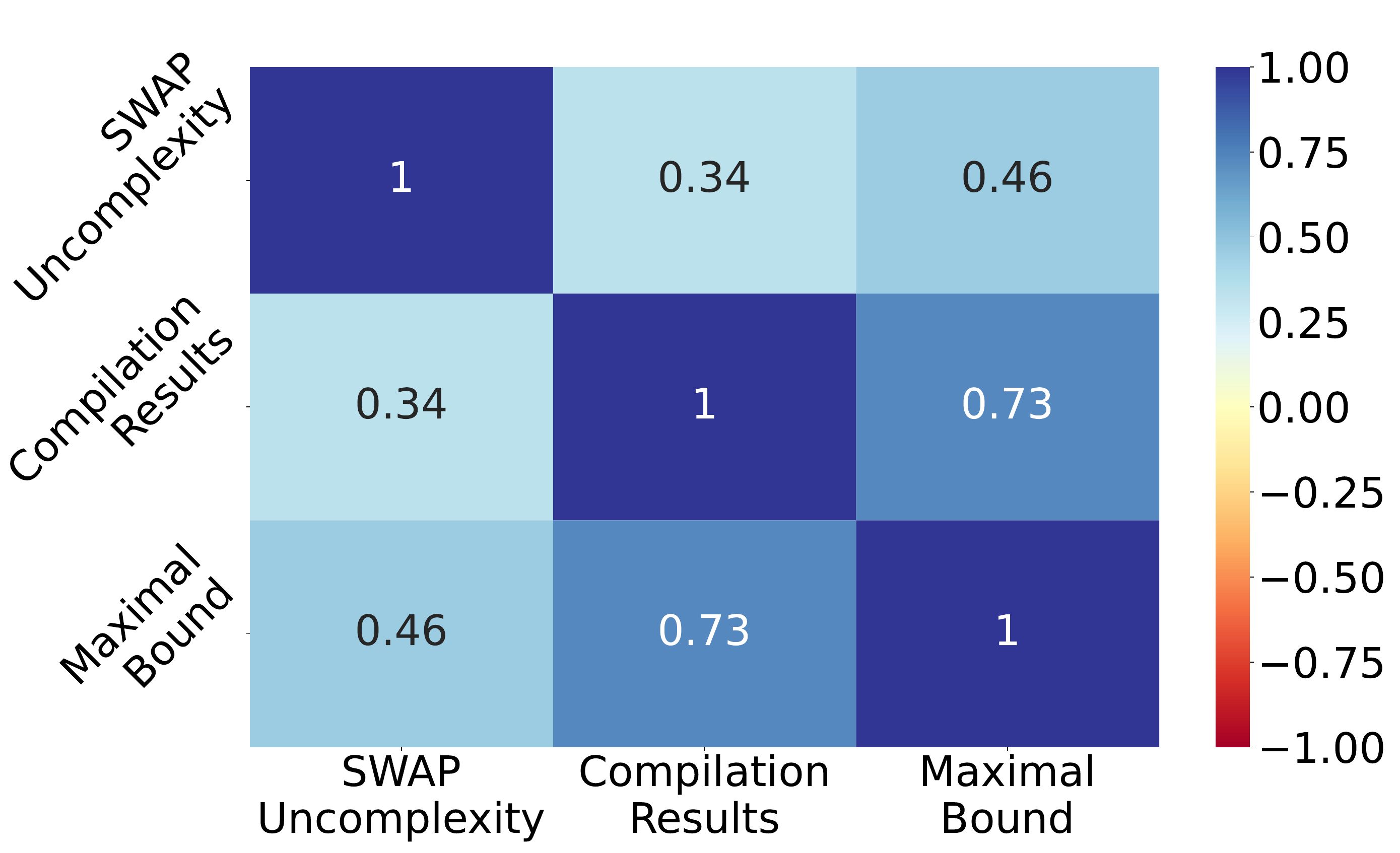}
\caption{The Pearson correlation matrix \cite{freedman2007statistics} between the critical parameters measured for our benchmark investigation. The values range between $-1$ and $1$ for negative and positive correlation, respectively. When one of the parameters changes, the other one changes in the same direction. In this figure, we observe a high positive correlation between all the selected parameters where the Pearson correlation coefficient ranges between 0.34 and 1.0.}
\label{figure:correlation}
\end{figure}

\begin{figure}
\centering
\includegraphics[width=\columnwidth]{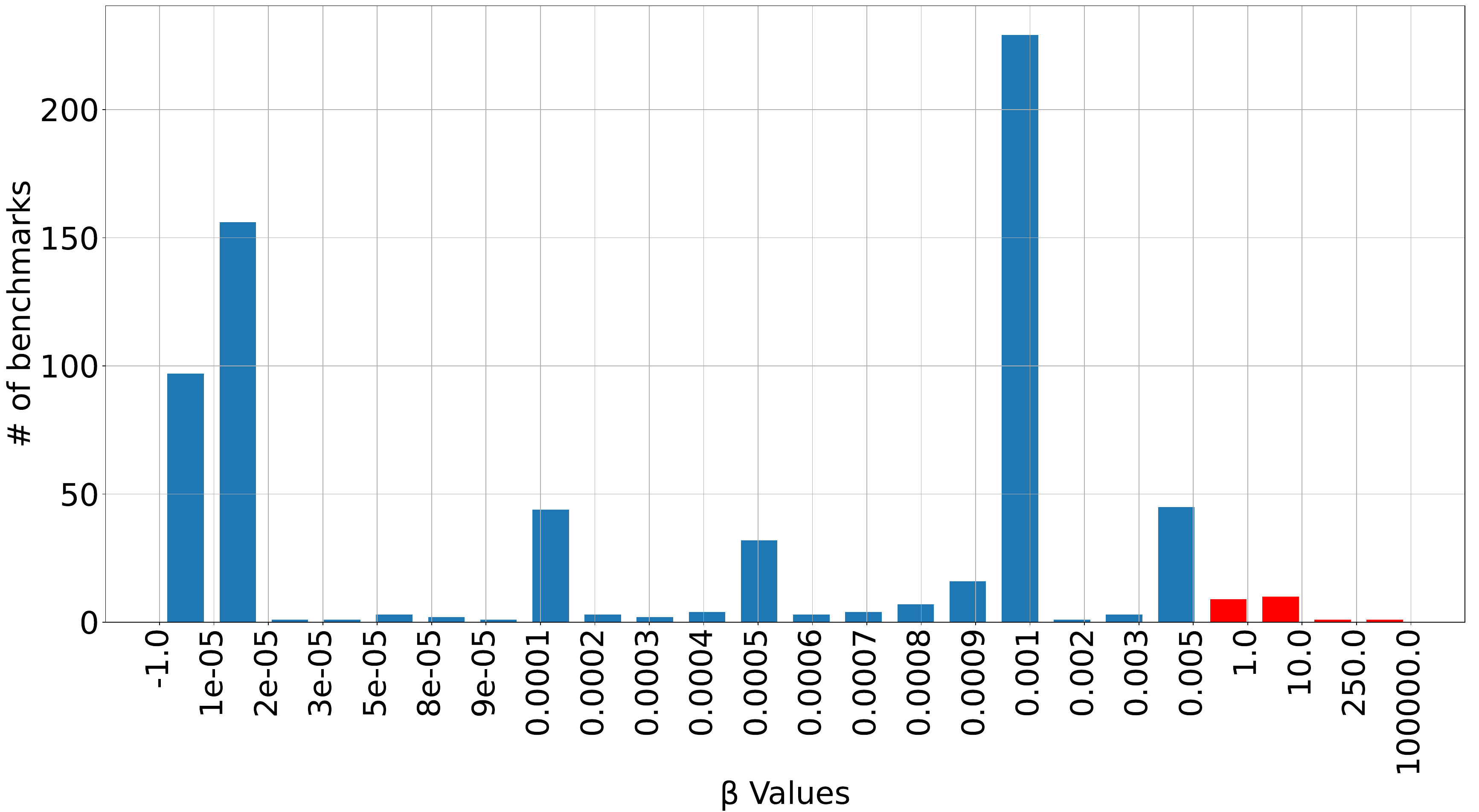}
\caption{The distribution of $\beta$ values for minimal bounds found by the SWAP uncomplexity algorithm. Results in blue ($\sim 97\%$) were found in the high-temperature regime described in \cite{PRE_beta_article}. We searched values in the range from $10^-5$ - $10^5$ based on the following formula $A*10^a$ where $A\in{1,2,3...9}$ and $a\in{-5,-4,...,4,5}$.}
\label{figure:betadistr}
\end{figure}

The results of our simulations are shown in \cref{figure:swapbounds,fig:swapbounds_bristlecone,figure:correlation,figure:betadistr}. In \cref{figure:swapbounds}, we display the normalized number of SWAP gates found by: the SWAP uncomplexity from \cref{section:thermo_path_length_uncomplexity_derivation} (shown in green); the Qiskit compiler (denoted in orange); and the maximum SWAP-gate bound (depicted in blue). We observe clearly that the SWAP uncomplexity can be reached but never surpassed by the Qiskit compiler for select benchmark trials. As expected, the Qiskit compiler significantly outperforms the maximum SWAP-gate count calculated. \cref{figure:swapbounds} also showcases the relation of the two-qubit gate count of the circuits (before compilation) and the bounds. The results in this figure, however, do not only depend on the circuit complexity, but also on the coupling graph. 

In order to more thoroughly scrutinize our results, we have included the relative graph-theoretic edge complexity for the benchmark circuits and have depicted them in \cref{fig:swapbounds_bristlecone}, Here, results are plotted for only one device, the Google Bristlecone device, and for a circuit size of up to six qubits. Two subgraphs are shown, relating the normalized number of SWAP gates to two different measures: in a) the relation between the bounds and the number of two-qubit gates and in b) the relation between the bounds and the IG size of the circuit, shown as nodes-edges pairs with correspondingly small IG figures as guides for the reader. It is evident that, while the number of IG nodes or qubits has the biggest influence on the results, the number of edges and gates is also important. We refer the reader to \cref{tab:benchmarks} in order to locate the circuit corresponding to points labeled on either of the horizontal axes of \cref{fig:swapbounds_bristlecone}. 

The non-triviality of the maximal and minimal SWAP-gate counts becomes evident in \cref{figure:correlation}, where we present a covariance matrix with correlation coefficients ranging as $[-1,1]$, with $0$ indicating no correlation \cite{freedman2007statistics}. This matrix compares the results that we obtained throughout the simulation; in particular, we compare the effective correlation between the SWAP uncomplexity; the Qiskit compiler SWAP calculation results; and the maximal bound as calculated in \cref{section:maxboundcalc}. Notably, both of our bounds (i.e., those obtained from our minimal bound with the SWAP uncomplexity algorithm, as well as the maximum SWAP-gate count) exhibit a substantial positive correlation ($34\%$ and $73\%$, respectively) with the actual results obtained from the compiler. The correlations here exemplify the non-triviality of the bounds; in other words, the SWAP uncomplexity and maximal bounds grow proportionally with the actual compilation results. The results at best coincide with each other, meaning that the lower bound equals the actual SWAP-gate count from the presence of a graph isomorphism; in this case, the SWAP uncomplexity, Qiskit result, and the maximal bound all obtain the same amount (which is zero if a graph isomorphism is present). These checks provide not only hard evidence for the usability of our methods, but additionally serve as a crucial sanity test that was passed for the algorithmic realization of the SWAP uncomplexity.

Although not shown in \cref{figure:correlation}, it is also worth observing the considerable impact of the initial placement on the resulting bounds; this particularly depends on the GED and the number of missing edges in the chosen CG partition compared to the IG. We therefore calculated the correlation coefficients of these two parameters (as well as the case when compared to our retrieved bounds), resulting in correlations of $79\%$ and $61\%$ for the SWAP uncomplexity and maximal bounds, respectively. Using the same initial placement for the Qiskit compiler resulted in a $45\%$ correlation with the parameters related to initial placement. 

The qubit-assignment strategy was initially tested with $729$ benchmarks, showing a success rate of $92.6\% $. The remaining $7.4\%$ of the benchmarks could not be finished due to insufficient computing resources. Recognizing the limited scalability of the approach (up to $16$ circuit qubits), we developed a more relaxed method for complete graphs, mentioned in \cref{section:qubit_assignment}. Indeed, the scalability of our exact algorithm already exceeded that of the exact state-of-the-art algorithms, which struggled beyond $6$ qubits \cite{siraichi2018qubit}. Furthermore, our initial placement encountered no difficulties in exploring a vast search space. It successfully executed circuits on all tested devices, extending up to a size of $72$ physical qubits in our case.

\cref{figure:betadistr} displays the $\beta$ values obtained during the course of the simulation. As we must search over a range in order to find the most-appropriate $\beta$, it is helpful and interesting to catalog roughly how many benchmarks exhibited the minimal bound obtained and at which $\beta$ values. In particular, we find that the vast majority of the benchmark pairs ($\sim 97\%$) led to minimal SWAP counts within the range of $\sim 10^{-5} - 10^{-3}$, consistent with the \emph{high-temperature regime} studied in \cite{PRE_beta_article}. We will comment on this more in \cref{section:discussion}.

\begin{figure}
\centering
\includegraphics[width=0.9\columnwidth]{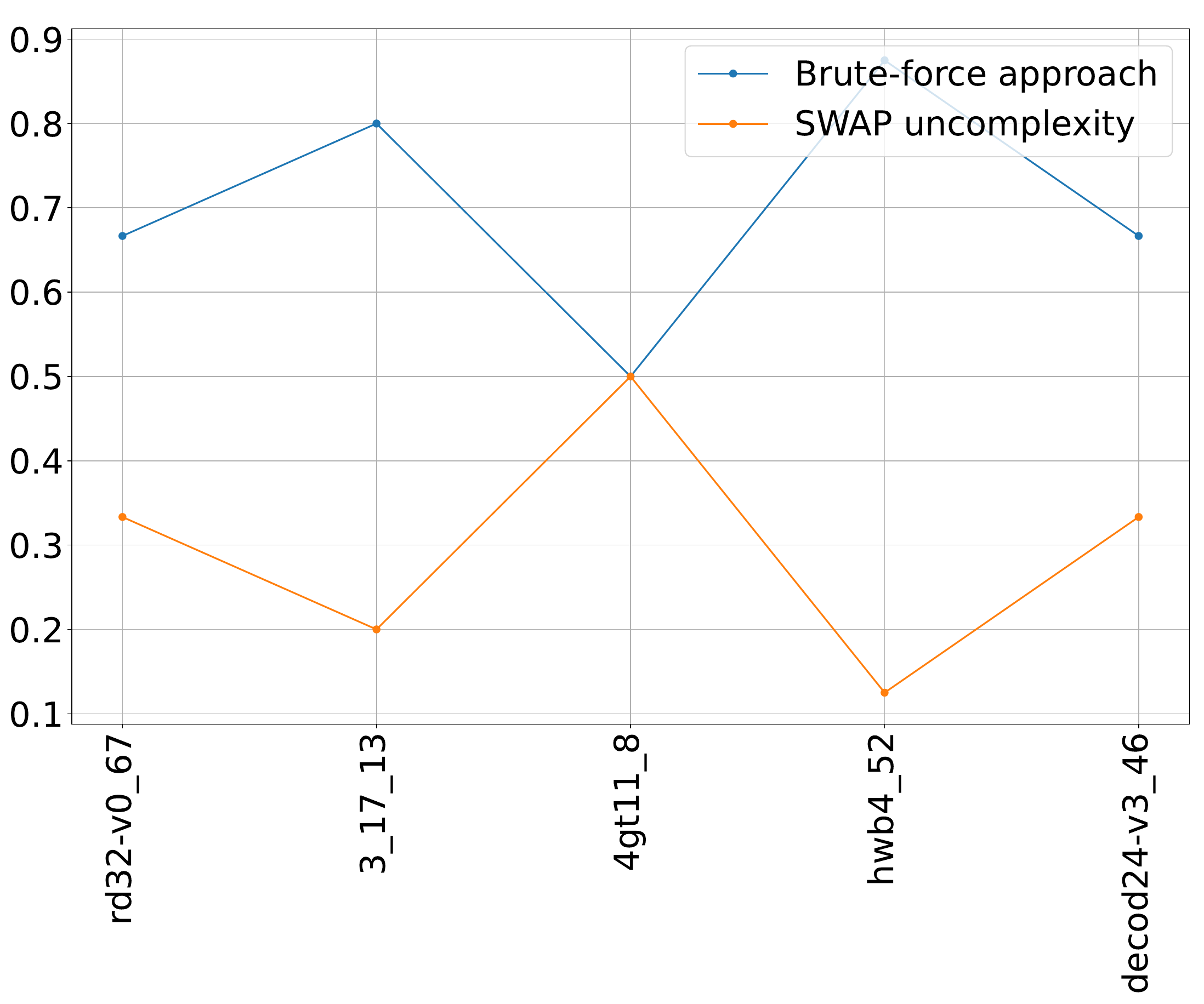}
\caption{Comparison of our minimal bounds with results of \cite{BF6qubits_wille}; the vertical axis charts the normalized number of SWAP gates. As before, the SWAP uncomplexity (shown in orange) is always smaller or equal to the actual number of SWAPs found by the brute-force compiler. The benchmarks employed in this study are identical to those utilized in \cite{BF6qubits_wille}. The results were normalized in the same way as in \cref{figure:swapbounds} .}
\label{figure:finalTest}
\end{figure}

Lastly, \cref{figure:finalTest} shows a comparison between our SWAP uncomplexity algorithm (\cref{section:algorithmic_implementation}) and the brute-force optimization results from \cite{BF6qubits_wille}. In this approach, the authors utilize an optimizer which essentially tries every permutation of SWAP placements possible while respecting the gate-dependency graph and weighted IG of the original quantum circuit. In all cases, we see clearly that the brute-force algorithm only achieves but never surpasses the SWAP uncomplexity bound.

\section{Discussion}\label{section:discussion}

It is known that the QCMP is NP-complete \cite{OnQubitMappingProblem}. As such, we have made three simplifications in order to derive the SWAP uncomplexity. Firstly, we do not consider single-qubit interactions, as it is known that such gates do not heavily affect calculated success rates \cite{qubitallocation}. Secondly, we have removed all two-qubit interaction noise from the CG; this should come as no surprise, as we are mainly interested in finding a lower bound for the number of SWAP gates required, and such a lower bound mandates the existence of a hypothetically noiseless quantum processor. Thirdly, we consider the limit in which gate dependencies for the IG are not considered; implicitly, we assume the existence of not only a noiseless quantum processor, but additionally one that can perform all two-qubit gate interactions required in one unit time slice, i.e. one whose two-qubit gate operations are infinitely parallelizable. 

\begin{figure}
\centering
\includegraphics[width=\columnwidth]{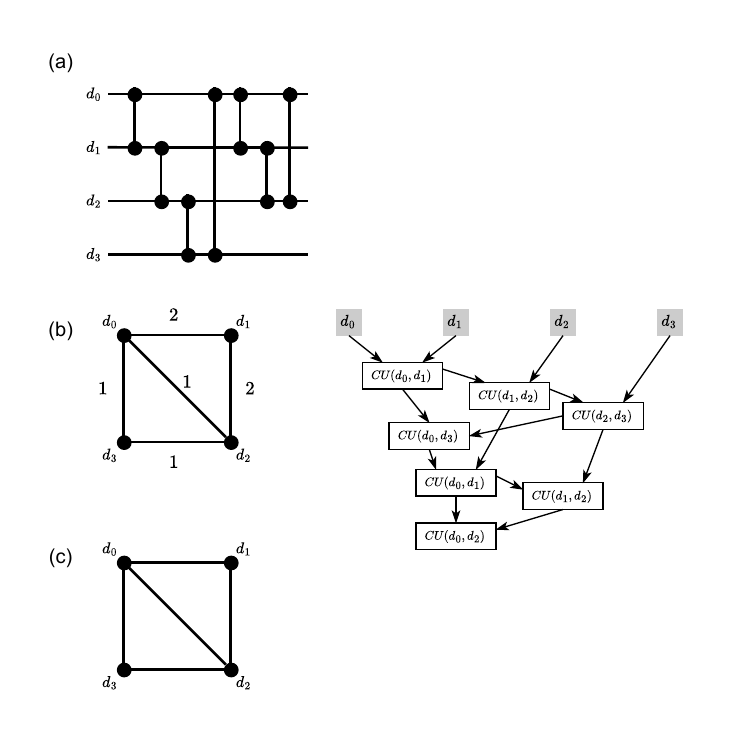}
\caption{Distinctions between the standard IG and the minimal version of the same IG. (a) illustrates an archetypal quantum circuit with arbitrary CU interactions, although making an exception with SWAP gates. (b) displays the standard weighted IG, together with its corresponding gate-dependency graph; we follow the gate-dependency notation used in \cite{lao2021timing}. (c) shows the unweighted form of our IG; here we do not take into account the gate ordering nor the number of gate calls realized per qubit, opting instead towards an unweighted IG.}
\label{fig:gatedependency_vs_uncomplex_IG}
\end{figure}

One may consider such a graph-theoretic interpretation of the SWAP uncomplexity as considering only connectivity dissimilarities between the interaction and coupling graphs. These differences are shown in \cref{fig:gatedependency_vs_uncomplex_IG}. In (a), we depict a generic quantum circuit, again omitting the single-qubit gates and taking every two-qubit gate shown to be a general CU gate for simplicity (the obvious exception to this rule would be the use of SWAP gates). In (b) the standard weighted IG and its accompanying gate-dependency graph are shown, following the notation of \cite{lao2021timing}. Finally, (c) makes manifest the differences between the standard IG for the QCMP and for our unweighted version. Here, we take any number of two-qubit interactions to $1$ on the IG, and we omit the gate-dependency graph, preferring an indefinite causal structure \cite{brukner2014quantum,goswami2020experimentsquantumcausality}. 

\begin{figure}
\centering
\includegraphics[width=1.05\columnwidth]{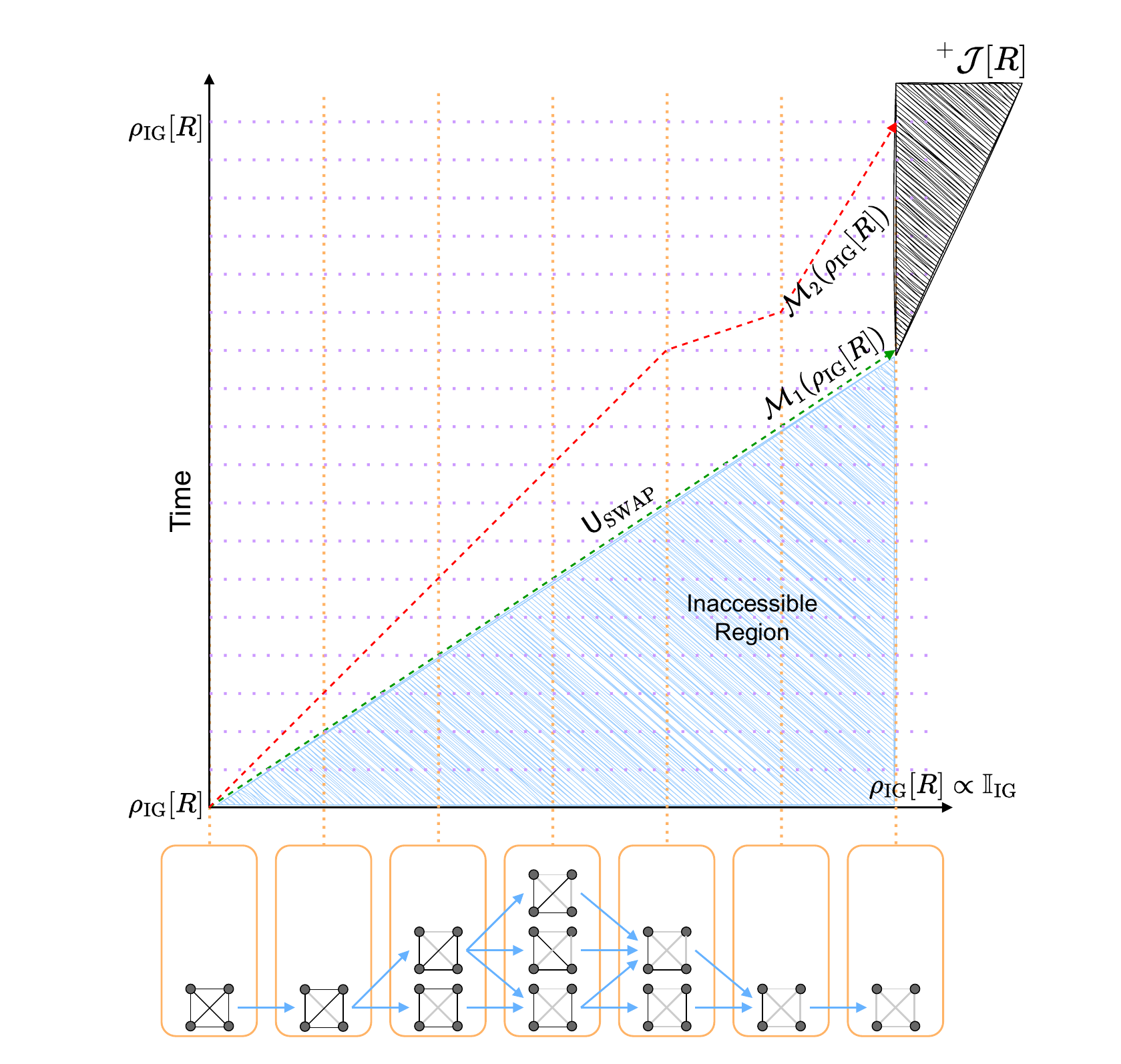}
\caption{A Penrose diagram representing the quantum circuit complexity for the evolution of a quantum state, which is related to a quantum circuit $\bm{\rho}_{\text{IG}}$ (for the sake of simplicity, we choose the state $\bm{\rho}_{\text{IG}}$ to be derived from the $K_{4}$ graph, but one can choose other examples). It is possible to generate different quantum complexities by adding different amounts and orderings of SWAP gates to the circuit as we approach the event horizon of a black hole (shown as a shaded triangle). Here, the connectivity limitations of the CG capture the role of the background spacetime geometry \cite{carroll2019spacetime}, as both determine the ease by which certain operations can be performed. The SWAP uncomplexity, as it ignores the effects of time ordering and the amount of qubit-qubit interaction present, can be associated with the null lightlike geodesic $^{+}\mathcal{J}[R]$, shown as a green dashed line. Below this arrow, the bottom-right part of the diagram represents possible states that are inaccessible to us, given the restrictions of the CG, as well as the operations available to us (SWAP gates, in our case).}
\label{fig:penrose_diagram_complexity}
\end{figure}

These concepts can be related using \emph{Penrose diagrams} \cite{penrosediagram1,carroll2019spacetime}, as shown in  \cref{fig:penrose_diagram_complexity}. A Penrose diagram typically shows the causal structure of events unfolding in a \emph{spacetime geometry} \cite{carroll2019spacetime}. In \cref{fig:penrose_diagram_complexity}, the horizontal axis refers to purely spatial evolutions, which in our case are shown by potential SWAP erasures. Each of the sets outlined in sky blue represent elements of the same total number of edges, but different spectral properties. Each graph within a given set represents a unique spectral signature which can be shared by multiple four-node subgraphs. The vertical axis depicts the evolution of time, and is known as the \emph{null time geodesic}, under which the set of \emph{trivial time-ordered gate operations} (i.e. idling, which in our simplified picture, is noiseless) evolve $\bm{\rho}_{\text{IG}}$ from point $R$ to the same later state. Here, the trivial minimal SWAP-gate count for the QCMP is represented, under which no SWAP gates are ever applied in order to adapt the quantum circuit to the device and its connectivity restrictions; in effect, the state is left to freely evolve for an infinite amount of time, with no regards to gate operations.

Conversely, the red dashed line represents possible spacetime evolutions arising from the application of distinct instances of time-ordered quantum operations $\mathcal{M}_{1}(\cdot), \mathcal{M}_{2}(\cdot)$ (which in this case are insertions of SWAP gates and subsequent erasures from the IG). In this way, every trajectory on the diagram can be associated with a given SWAP uncomplexity from a sequence of quantum operations. The possible endpoints of the quantum circuit are shown along the shaded triangle in dark gray, which represents $\bm{\rho}_{\text{IG}} = \mathbb{I}_{n}$, i.e. the state of maximal circuit complexity, a maximally mixed state. 

Furthermore, erasure transformations on the original quantum circuit proceed according to restrictions dictated by the background geometry (which in our present case is analogous to the CG connectivity). Evolution commences at a spacetime point $R$. The green dashed line traces out the \emph{lightlike null geodesic} (i.e. future lightcone) $^{+}\mathcal{J}[R]$. This geodesic signifies the SWAP uncomplexity, which is the path that the state takes under the minimal set of causally-indefinite operations such that we approach the maximally-mixed state in minimal time. As we solve for the SWAP uncomplexity, without consideration of time ordering and as dictated by the thermodynamic path length calculable via the Fisher information \cite{measure_thermo_length,path_length1,path_length2,path_length3}, we can interpret our bound as a sort of \emph{lightcone evolution} of our initial state $\bm{\rho}_{\text{IG}}$ towards the event horizon of a black hole (shown as the shaded triangle). Previous work has already alluded to the concept of optimization over thermodynamic distance \cite{path_length3,PhysRevLett.121.030605stochastic_thermo_length}; as such, our results point to a natural and reasonable extension of this trend for the quantum circuit mapping problem. 

As we touched upon earlier, it is possible to interpret the shaded triangle in \cref{fig:penrose_diagram_complexity} as the event horizon of a black hole. Consider a benign black hole scenario in which the black hole itself can only erase information from a density matrix in accordance with only certain SWAP gates from some constrained architecture (i.e., the black hole itself is described with respect to a background geometry, which constrains which operations can be performed). As black holes are known to be the fastest information scramblers in nature \cite{sekino2008fastsusskind}, the QCMP can be viewed through the lens of a scrambling process, yielding the most-efficient method to maximally mix the information of the IG's density matrix for a given $\beta$. This process exemplifies the traits of quantum circuit uncomplexity and we have shown that the quantity $\mathsf{U}_{\text{SWAP}}$ can in fact be optimized for using our technique.

Taking stock, we would then expect that any realistic quantum compiler which takes into account time ordering and finite qubit-qubit interactions per unit of time slicing to be limited by the lightlike null geodesic. Indeed, surpassing the lightlike null geodesic would introduce operations which are not inside of the lightcone, giving access to the uncomputable region to the bottom-right. Such trajectories could be made possible using a larger set of routing resources, such as teleportation \cite{teleportation_wille,gorshkov}. As a concrete counterexample, consider a hypothetical quantum compiler which could surpass $\mathsf{U}_{\text{SWAP}}$. One of the main assumptions that we utilize in our formulation above is that possible interactions can occur if and only if a subsystem interaction between qubits exists in the density matrix picture. Consequently, moving outside the region embellished by the future lightlike geodesic corresponds to new operations which must be taken into account. One simple example lies in \emph{teleportation-based} quantum-circuit mapping, which can be used to swap CG qubits which do not share a physical subsystem interaction, and can allow for a smaller quantum circuit complexity \cite{gorshkov}. From the standpoint of our formalism, this difference would correspond to allowing for non-nearest-neighbor permutation matrices to arise in the doubly-stochastic quantum channel described in \cref{eq:kraus_decomp}. One may suspect that architectures in the future may benefit from such on-chip teleportation procedures, as work has shown that speedups exist over classical SWAP methods for exchanging distant qubits \cite{gorshkov,teleportation_wille,devulapalli2022quantum}, albeit with larger circuit and entanglement overhead.

Finally, we close this section by discussing the $\beta$ parameter in more detail. For the vast majority of our results (approximately $\sim 97\%$, as shown in \cref{figure:betadistr}), the $\beta$ values yielding the SWAP uncomplexity for an IG/CG pair coincide with the high-temperature regime noted in \cite{PRE_beta_article}. Indeed, it was shown there that the high-temperature regime allows for optimization of the quantum relative entropy for density matrices contrived from complex networks. The main interpretation provided in \cite{PRE_beta_article} is that the parameter $\beta$ controls the diffusion of information about the graph neighborhood to other vertices. Conversely, if we take the $\beta \mapsto 0$ limit, diffusion over the vertices of the graph is limited, and therefore, only information about the degrees of links is conferred. This trait is known to exhibit a first-order linear dependency in the adjacency matrix, and as such, the tendency for information to diffuse over the network becomes uniform. As we can see from \cref{figure:betadistr}, less than $3\%$ of our benchmarks do not fall within the high-temperature $\beta$ range described in \cite{PRE_beta_article} (shown in red). Those outliers, however, can be explained with the triviality of the correct $\beta$ search process, which was particularly based on sweeping over a range of values between $10^{-5} - 10^{5}$. One future direction, therefore, is to devise more efficient schemes for finding the optimal $\beta$ value, guaranteeing a minimal bound in less computation time. One could imagine that such a goal could be completed by use of a gradient optimization method, but we leave such exploration to future work.

\section{Conclusions \& Outlook}\label{section:conclusion} 

The QCMP itself has been described using several approaches from computer science, many of which have allowed for the development of entirely new strategies for solving the problem. Our contribution here serves a different purpose. At the level of theoretical physics, as well as quantum information theory, we have solved a simplified subproblem of the QCMP which we dub the ``lightcone bound" to the QCMP. We have shown that solving for the lightcone SWAP uncomplexity bound is optimal, in the sense that it implicitly defines the shortest path through a configuration space of restricted gate operations. At precisely the SWAP uncomplexity limit, it is expected that: the quantum device is noiseless; the two-qubit gate interactions given by the device can be performed with an indefinite causal order; and that any number of two-qubit interactions can be performed in tandem within one unit time slice. Therefore, we surmise that no quantum compiler in existence can violate this lower bound, allowing for a fundamental means of comparison between differing strategies for solving the real-world QCMP. We have provided, to the best of our knowledge, the first instance of a solvable lower bound for SWAP-gate count in the context of quantum-circuit compilation. The SWAP uncomplexity was derived using tools from graph theory, quantum information theory, quantum circuit complexity, and information geometry. In addition to the use case discussed in this work, potential applications of uncomplexity for quantum machine learning are discussed in \cite{sarkar2022applications}. This work also represents the first application of quantum circuit uncomplexity to the realm of practical quantum information processing. 

Our original purpose for deriving the SWAP uncomplexity has been to inform and create a meaningful method of comparison between quantum compilers and strategies for solving the practically-motivated QCMP. This goal has been accomplished, as the SWAP uncomplexity sheds light onto the physical nature of the QCMP as a lower bound for thermodynamic path length under restricted gate transformations. Therefore, the use of SWAP uncomplexity as a metric for routing efficiency is pragmatic and justified, and can be used to quantify and compare routing strategies, as well as helping to inform architectural decisions by quantum architects and processor designers. 

Of independent interest may be the qubit-assignment algorithm which was designed to aid in the calculation of the SWAP uncomplexity. This algorithm, grounded as a graph similarity search, inspects distinct $n$-qubit partitions of a given CG, and returns the most-similar resultant to the IG provided. Employing this method has enabled us to map circuits with up to $16$ qubits onto devices with up to $72$ physical qubits. For larger circuits, we devised an alternative approach. The initial placement precedes the minimal SWAP-gate count solution, which is further utilized for routing and minimal bound calculation. Additionally, we calculated a maximal bound by leveraging known classical graph metrics; both of these novel structures provide additional tools of interest outside of the scope of this work. 

We would now like to draw attention to several open problems regarding our work, as well as several future possible directions for research:

\begin{enumerate}[nolistsep,noitemsep]
\item \emph{Improvements to the subgraph similarity search algorithm.} In this work, although our qubit-assignment algorithm outperforms the current state of the art solver \cite{qubitplacement_for_wille_6qubits}, we were still limited by the scalability of the qubit-assignment algorithm constructed in \cref{section:qubit_assignment}. However, once a suitable qubit assignment is set, the calculation of $\mathsf{U}_{\text{SWAP}}$ can be completed in $\sim \mathcal{O}(\text{dim}(\bm{\rho}_{\text{CG}})^{4})$ timesteps, as per the Birkhoff-Von-Neumann algorithm \cite{johnson1960BVNalgorithm}. A future research goal could involve making further scalability improvements to the qubit-assignment algorithm, or by considering more advanced methods of routing, such as those using \emph{ancillary qubits} \cite{lao2021timing}. 
\item \emph{Searching for the optimal $\beta$ value.} In our work, we have taken a somewhat naive approach to optimizing for $\beta$; however, because of the similarity of \cref{figure:betaVSvne} to a phase diagram, one may be able to use concepts from condensed matter theory \cite{goldenfeld2018lectures,brightwell1999graph,phase_transitions_complex_networks} in order to devise a suitable gradient-based optimization method.
\item \emph{Analytical expression for tightness of the SWAP uncomplexity to the brute-force solution.} We have given empirical evidence for tightness, but it still remains to define an analytical expression for how similar in general our solution is, compared to the brute-force solution proposed in \cite{BF6qubits_wille}, and how tightness scales as the size of the quantum circuit to be mapped increases in both register and depth.
\item \emph{Extension to incorporate bridge gates, teleportation-based quantum circuit mapping, and shuttling.} There are other methods commonly in use, in addition to the SWAP gate, for conforming a quantum circuit to hardware. Our approach is extendable for the Bridge gates mentioned in \cite{bridge1,itoko2020optimization,bridge2,liu2023tacklingbridge3,bridge4}, as well as the quantum teleportation-based protocols of \cite{teleportation_wille,gorshkov,hillmich2021exploiting,devulapalli2022quantum} and shuttling-based approaches for spin-qubit architectures \cite{nikiforos_besnake,nikiforos_spinq}, trapped-ion architectures \cite{trapped_ion_1,sivarajah2020t}, and neutral-atom devices \cite{patel2022geyser}.
\item \emph{Extension for quantum error correction codes, in particular syndrome extraction circuits.} It is well-known that various \emph{fault-tolerance} protocols are required in order to ensure that quantum error correction codes function up to their full code distance \cite{preskill1998reliable,qec1,gottesman2010introductionqec2,PhysRevA.57.127qec3,chamberland2018flagqec4,PhysRevA.68.042322qec5,cross2007comparativeqec6,campbell2017roadsqec7,aliferis2005quantumqec8,aharonov1997faultqec9,knill2005quantumqec10,PhysRevLett.121.050502qec11}. As our bound constitutes a non-trivial resource requirement, it may be useful to adapt fault-tolerance protocols further to the setting of quantum compilation, in which an error correction code is adapted to a device not specifically designed for a particular code family \cite{lidar2013quantumqec12,PhysRevA.101.032333qec13}. 
\item \emph{Extension for entanglement/qubit routing in quantum communications networks and modular architectures.} Several other extensions may be possible as well, including those allowing for bounds on the QCMP for modular scenarios \cite{bandic2023mapping_mod1} as well as for \emph{entanglement distribution} in noisy quantum networks \cite{pant2019routing1}.
\end{enumerate}

Finally, we remark that the problem of assessing similarities between two complex networks is a problem spanning many disciplines. Indeed, our work follows recent trends of utilizing quantum information theory and statistical mechanics to study complex networks \cite{biamonte_nature,biamonte_spectral,asymptoticentropy_diffusiontimepaper,PRE_beta_article}. As the task of comparing the distance between graphs appears in many different areas of science \cite{strogatz2001exploring,statmech_complexnetworks}, we expect the implications of our work to stretch beyond the realm of quantum information science.

\section{Software Availability}

The software developed for this project is available at \noindent\href{https://github.com/QML-Group/QCMP-complexity-bound}{https://github.com/QML-Group/QCMP-complexity-bound}.

\section{Acknowledgements}

We thank Kenneth Goodenough, Hans van Someren, Pablo le Henaff,  Luise Prielinger, David Elkouss, and Tariq Bontekoe for insightful discussions and useful manuscript feedback. MS, MB, and SF are grateful for financial support from the Intel corporation. CGA also acknowledges support from the Spanish Ministry of Science, Innovation and Universities through the Beatriz Galindo program 2020 (BG20-00023) and the European ERDF under grant PID2021-123627OB-C51. Also from the QuantERA grant EQUIP with the grant number PCI2022-133004, funded by Agencia Estatal de Investigación, Ministerio de Ciencia e Innovación, Gobierno de España,MCIN/AEI/10.13039/501100011033, and by the European Union “NextGenerationEU/PRTR. AS acknowledges funding from the Dutch Research Council (NWO) through the project ``QuTech Part III Application-based research" (project no. 601.QT.001 Part III-C—NISQ).

\section{Author Contributions}

MS, AS, and SS developed the theoretical framework and formalism. MB, SS, and AS developed the numerical algorithm based on the framework and implemented all numerical simulations. MB developed the subgraph isomorphism qubit-assignment algorithm. Part of this work was conducted during the master thesis for SS, and was supervised by MS, MB, and SF. CA and SF supervised and coordinated the project, and provided guidance during the writing process.

\bibliography{bibliography}


\clearpage
\appendix
\section{Graph Theory} \label{section:GraphTheoryBackground}

In \cref{figure:examplemapping}, the IG and CG are examples of \emph{simple, undirected graphs} \cite{graph_th_book}, with \emph{simple} referring to a restriction of only one edge between any two vertices, and \emph{undirected} meaning without directionality indicated by arrows. For this work, we restrict ourselves to this regime only. An example of a simple, undirected graph can be seen in \cref{figure:examplemapping}B. Here, we define a graph as an ordered-pair object $G(E_{G},V_{G})$ with edge set $E_{G} = \{ e_{ij} | i,j \in V_{G}, i \neq j\}$ and vertex set $V_{G}$ containing all nodes of $G$ \cite{algebraic_godsil_graph_theory}. Additionally, we define the \emph{adjacency matrix} $\mathbf{A}$, \emph{degree matrix} $\mathbf{D}$, and \emph{combinatorial graph Laplacian} $\mathbf{L}$ from literature \cite{biamonte_spectral,biamonte_nature}. As usual, the graph Laplacian takes on the form $\mathbf{L} = \mathbf{D} - \mathbf{A}$.

From here onwards, we will refer to the combinatorial graph Laplacian as simply the \emph{graph Laplacian}. Several forms of the graph Laplacian exist in the literature \cite{algebraic_godsil_graph_theory,graph_th_book,biamonte_nature,biamonte_spectral,severini,asymptoticentropy_diffusiontimepaper}; as such, we have opted to use the definition of the graph Laplacian which upholds the triangle inequality \cite{biamonte_spectral,biamonte_nature}. The graph Laplacian in the current context is known to be symmetric and positive semidefinite. 

Let us also define the notions of \emph{graph isomorphism} and \emph{homomorphisms}, as we will use these later. For two graphs $G(E_{G},V_{G})$ and $H(E_{H},V_{H})$ with $|V_{G}| = |V_{H}|$, we say that $G$ and $H$ are \emph{cospectral} if they share the same eigenvalue spectrum \cite{algebraic_godsil_graph_theory}. Note that the graph Laplacian's eigenvalue spectrum is \emph{not sufficient} in order to determine whether or not two graphs are the same; this reason motivates us towards the treatment described in \cref{section:graphs_as_density_matrices} and originally treated in \cite{biamonte_nature,biamonte_spectral} with Gibbs states.

Additionally, we say that $G$ and $H$ are \emph{graph homomorphic} if there exists a map $G \mapsto H$ such that the vertex and edge connectivities of the two graphs are preserved; we represent this relation in the text as $G \triangleq H$. The task of determining whether or not two graphs exhibit a graph isomorphism or homomorphism is in general NP-complete \cite{algebraic_godsil_graph_theory,graph_th_book}, although certain exceptions exist. If we are looking for an \emph{embedding} of graph $G$ onto graph $H$ such that, for some subgraph $H' \subseteq H$, we have $G \triangleq H'$. The generalization of this problem is known as the \emph{subgraph-isomorphism problem} (SIP). As we allow $|V_{G}| \subseteq |V_{H}|$, the task adds an additional layer of complexity, since we must identify a suitable subgraph for comparison; this subgraph is known as an \emph{induced subgraph}, and is defined as a subgraph $G' \subseteq G$ for which $V_{G'} \subseteq V_{G}$ and $E_{G'} \subseteq E_{G}$.

We further remark that solving the SIP efficiently is still an active area of research, and various approaches exist for addressing it. These methods range from brute-force approaches to more sophisticated algorithms that exploit specific properties of the graphs being analyzed. Some of the most common algorithms used for solving the SIP include the Vento-Foggia algorithm (VF2) \cite{cordella2004sub, sansone2001improved}, RI \cite{bonnici2013subgraph}, The LAD (Labeled Anatomy Directed) \cite{solnon2010alldifferent} and Ullman \cite{ullmann2011bit} algorithms. Each algorithm has its strengths and weaknesses, and the choice depends on the specific requirements of the application. \emph{VF2 algorithm} is a most commonly used algorithm for solving the SIP known for its speed and effectiveness. It works by maintaining a partial matching between the nodes of the pattern and target graphs. It starts with an empty matching and gradually extends the matching by adding pairs of nodes, one from the pattern graph and one from the target graph, until a complete matching is found or it is determined that no matching exists. It has a worst-case time complexity of $O(|V|^2 * |E|)$, where $|V|$ is the number of nodes in the graph and $|E|$ is the number of edges.

If an exact isomorphism cannot be found, one must choose a subgraph known to be close to a graph isomorphism, i.e., as close to a graph homomorphism as possible. Just as for the SIP, many approaches exist for extending into the regime of subgraphs \cite{koutra2011algorithms,zager2008graph,samanvi2015subgraph} exhibiting graph homomorphisms. Typically, one employs some graph-theoretic distance metric in order to locate the most similar subgraph \cite{roth2006introduction,abu2015exact, sellers1974theory}.

Finally, as we will require the use of the \emph{graph-edit distance} (GED) in \cref{section:qubit_assignment}, we give a brief definition of this metric. The GED itself is known as a \emph{classical similarity measure} between two graphs $G$ and $H$ \cite{graph_th_book,algebraic_godsil_graph_theory}. Given a set of \emph{graph-edit operations} (such as edge addition or edge removal in our case), one may define the GED as:
\begin{equation}
\text{GED}(G,H) = \min_{\{ e_{1} \dots e_{i}\} \in S_{\text{ops}(G,H)}} \sum_{i} c(e_{i}),
\end{equation}
where $\{e_{i}\} \in S_{\text{ops}(G,H)}$ represents the set of graph-edit operations along all possible graph-edit paths between graphs $G$ and $H$. $c(e_{i})$ is the \emph{cost} of the graph-edit operation. An exact algorithmic implementation of the GED usually can be found in the \emph{A$^{*}$ search algorithm}, wherein the problem of finding the minimal graph-edit cost is transformed into a \emph{shortest-path algorithm} \cite{nocedal1999numerical,russell2010artificial}. However, in this work, we introduce an algorithm based on the depth-first GED (\emph{DF-GED})  algorithm \cite{abu2015exact}, which we detail in \cref{section:qubit_assignment}.

\section{Quantum Information Theory} \label{section:quantum_information_theory_basics}

In this section, we review some of the basics of quantum information theory which may be useful later. We refer the reader to \cite{wilde_qit,nielsenchuang,watrous2018theory} for a more nuanced treatise of quantum information theory. 

In quantum computation and quantum information theory, the fundamental information unit is known as a \emph{qubit}. Any pure qubit state can represented as:

\begin{equation}
\ket{\psi} = \alpha\ket{0} + \beta\ket{1}~,
\end{equation}

where $|\alpha|^{2} + |\beta|^{2} = 1$. Qubits are defined in a \emph{Hilbert space} $\mathcal{H}$; these are known as \emph{inner-product spaces} on the field of complex numbers. 

Representing systems of many qubits becomes a cumbersome task when such systems are entangled with other systems. For pure quantum states, it suffices to utilize Dirac notation, but, when mixed states are involved, one conventionally uses the language of \emph{density matrices} $\bm{\rho}$, whose explicit form is given by:

\begin{equation}
\bm{\rho} = \sum_{j} p_{j} \ket{\psi}\bra{\psi}~,
\end{equation}

where $p_{j}$ represents the probability of each pure state in the ensemble. 

Density matrices exhibit several important properties for the present work. Density matrices in general are:

\begin{itemize}[nolistsep,noitemsep]
\item Defined formally as objects $\bm{\rho}$ such that $\bm{\rho} \in \mathcal{M}_{d}(\mathbb{C}) \cong \mathcal{B}(\mathcal{H}^{d})$, equipped with a Hilbert-Schmidt scalar product as usual (where $d$ is the dimension of the Hilbert space). Here, $\mathcal{M}_{d}(\mathbb{C})$ represents the set of complex-valued $d \times d$ square matrices, and $\mathcal{B}(\mathcal{H}^{d})$ is the set of bounded linear operators on a Hilbert space \cite{wolf_qchannels};
\item Normalized, such that $\Tr(\ket{\psi}\bra{\psi}) = 1$;
\item Positive, such that $\bm{\rho} \geq 0$;
\item Hermitian, meaning $\bm{\rho} = \bm{\rho}^{\dagger}$; and 
\item Projectors, where $\bm{\rho} = \bm{\rho}^{2}$.
\end{itemize}

\section{Calculating the Maximal Bound} \label{section:maxboundcalc}

We briefly introduce here a metric related to qubit assignment in order to approximate a maximal bound. This bound was employed in \cref{section:benchmark_evaluation_and_results}. The metric $\mathcal{M}_{\text{SWAP}}$ involves assessing the \emph{diameter} $\mathds{D}(\cdot)$ \cite{hernandez2011classification} of the CG subgraph $G'_{\text{CG}}$ that is selected for qubit assignment. We then simply multiply the diameter with the amount of two-qubit gates, equivalent to the number of edges, $|E_{\text{IG}}|$ in the circuit. This forms our approximate maximal bound for the number of SWAP gates, and takes on the following algebraic form:

\begin{equation}
\mathcal{M}_{\text{SWAP}}(E_{\text{IG}},G'_{\text{CG}}) = |E_{\text{IG}}| \cdot \big( \mathds{D}(G'_{\text{CG}})-1 \big)~.
\end{equation}

By considering the maximal distance between any two points in the chosen CG subgraph and imagining the worst-case scenario in which all the two-qubit gates are on the path between those two points, we get this maximal bound. This approach, although approximate, offers a broader perspective on the possible SWAP overhead associated with the qubit assignment, enhancing the depth and utility of our analysis. It is worth noting that, unlike the minimal bound derived in \cref{section:thermo_path_length_uncomplexity_derivation}, our approach for the maximal bound involves using \emph{weighted} IGs, taking into account the two-qubit gate depth of the underlying circuit; this makes our maximal bound specific to each benchmark.

\section{Benchmarks and quantum devices used for experiments}
\label{section:appendix1}

\begin{table}
\centering\footnotesize
\caption{Quantum devices used for simulations: We chose 16 of the most renowned device layouts of superconducting technology. We opted for superconducting qubits as the limit in connectivity characterizes them. The CGs range in size from 5 to 72 qubits.}
\begin{tabular}{|l|l|}
\hline
\textbf{Quantum device} & \textbf{Number of qubits} \\ \hline
IBM Athens & 5 \\ \hline
QuTech Starmon-5 & 5 \\ \hline
IBM Yorktown & 5 \\ \hline
IBM Ourense & 5 \\ \hline
QuTech Surface-7 & 7 \\ \hline
IBM Casablanca & 7 \\ \hline
Rigetti Agave & 8 \\ \hline
IBM Melbourne & 15 \\ \hline
Rigetti Aspen-1 & 16 \\ \hline
QuTech Surface-17 & 17 \\ \hline
IBM Singapore & 20 \\ \hline
IBM Johannesburg & 20 \\ \hline
IBM Tokyo & 20 \\ \hline
IBM Paris  & 27 \\ \hline
IBM Rochester & 53 \\ \hline
Google Bristlecone & 72 \\ \hline
\end{tabular}
\label{tab:devices}
\end{table}

\clearpage

\begin{table*}
\centering
\footnotesize
\caption{Benchmarks used for experiments taken from \cite{qbench2023}. The benchmarks are characterized by different IG connectivities and range from $3$ to $20$ qubits.}
\begin{tabular}{|p{3.5cm}|p{3.5cm}|p{3.5cm}|p{3.5cm}|p{3cm}|}
\hline
\textbf{Benchmark} & \textbf{Number of qubits} & \textbf{Number of gates} & \textbf{2-qubit gate \%} &\textbf{IG(nodes,edges)}\\ \hline
basis\_change\_n3 & 3 & 79 & 0,126582278 & (3,2)\\ \hline
fredkin\_n3 & 3 & 51 & 0,156862745  & (3,3)\\\hline
grover\_n3 & 3 & 102 & 0,117647059 & (3,3) \\ \hline
teleportation\_n3 & 3 & 20 & 0,1 & (3,2)\\ \hline
adder\_n4 & 4 & 63 & 0,158730159 & (4,4)\\ \hline
bell\_n4 & 4 & 66 & 0,106060606 & (4,3)\\ \hline
cuccaroAdder\_1b & 4 & 83 & 0,204819277 & (4,4)\\ \hline
q=4\_s=19996\_2qbf=02\_1 & 4 & 20000 & 0,20365 & (4,6)\\ \hline
q=4\_s=2996\_2qbf=08\_1 & 4 & 3000 & 0,814 & (4,6)\\ \hline
variational\_n4 & 4 & 94 & 0,170212766 & (4,3)\\ \hline
vbeAdder\_1b & 4 & 74 & 0,189189189 & (4,5)\\ \hline
4gt10-v1\_81 & 5 & 424 & 0,155660377 & (5,10)\\ \hline
4gt13\_92 & 5 & 190 & 0,157894737 & (5,6)\\ \hline
4gt5\_75 & 5 & 239 & 0,158995816 & (5,9)\\ \hline
alu-v1\_28 & 5 & 105 & 0,171428571 & (5,7)\\ \hline
alu-v2\_31 & 5 & 1295 & 0,152895753 & (5,9)\\ \hline
decod24-v1\_41 & 5 & 241 & 0,157676349 & (5,8)\\ \hline
error\_correctiond3\_n5 & 5 & 278 & 0,176258993 & (5,5)\\ \hline
q=5\_s=2995\_2qbf=09\_1 & 5 & 3000 & 0,899 & (5,10)\\ \hline
qec\_en\_n5 & 5 & 61 & 0,163934426 & (5,4)\\ \hline
qec\_sm\_n5 & 5 & 61 & 0,163934426 & (5,4)\\ \hline
quantum\_volume\_n5 & 5 & 411 & 0,124087591 & (5,8)\\ \hline
simon\_n6 & 5 & 92 & 0,152173913 & (5,5)\\ \hline
4gt12-v0\_87 & 6 & 711 & 0,157524613 & (6,12)\\ \hline
4gt4-v0\_72 & 6 & 740 & 0,152702703 & (6,12)\\ \hline
alu-v2\_30 & 6 & 1446 & 0,154218534 & (6,14)\\ \hline
ex3\_229 & 6 & 1153 & 0,151777971 & (6,10)\\ \hline
graycode6\_47 & 6 & 15 & 0,333333333 & (6,5)\\ \hline
mod5adder\_127 & 6 & 1577 & 0,151553583 & (6,13)\\ \hline
q=6\_s=2994\_2qbf=08\_1 & 6 & 3000 & 0,802 & (6,14)\\ \hline
q=6\_s=54\_2qbf=022\_1 & 6 & 60 & 0,233333333 & (6,8)\\ \hline
qaoa\_n6 & 6 & 528 & 0,102272727 & (6,9)\\ \hline
sf\_274 & 6 & 2221 & 0,151283206 & (6,11)\\ \hline
xor5\_254 & 6 & 17 & 0,294117647 & (6,5)\\ \hline
4mod5-bdd\_287 & 7 & 196 & 0,158163265 & (7,11)\\ \hline
alu-bdd\_288 & 7 & 240 & 0,158333333 & (7,12)\\ \hline
C17\_204 & 7 & 1341 & 0,152870992 & (7,20)\\ \hline
ham7\_104 & 7 & 922 & 0,161605206 & (7,16)\\ \hline
majority\_239 & 7 & 1754 & 0,152223489 & (7,16)\\ \hline
q=7\_s=2993\_2qbf=08\_1 & 7 & 3000 & 0,795333333 & (7,21)\\ \hline
q=7\_s=29993\_2qbf=08\_1 & 7 & 30000 & 0,799866667 & (7,21)\\ \hline
dnn\_n8 & 8 & 1904 & 0,100840336 & (8,8)\\ \hline
f2\_232 & 8 & 3456 & 0,151909722 & (8,22)\\ \hline
hwb7\_59 & 8 & 70093 & 0,152383262 & (8,28)\\ \hline
q=8\_s=2992\_2qbf=01\_1 & 8 & 3000 & 0,091666667 & (8,28)\\ \hline
vqe\_uccsd\_n8 & 8 & 24136 & 0,22737819 & (8,19)\\ \hline
q=9\_s=19991\_2qbf=08\_1 & 9 & 20000 & 0,79645 & (9,36) \\ \hline
q=9\_s=2991\_2qbf=01\_1 & 9 & 3000 & 0,101 & (9,36)\\ \hline
q=9\_s=51\_2qbf=012\_1 & 9 & 60 & 0,116666667 & (9,6)\\ \hline
adder\_n10 & 10 & 328 & 0,198170732 & (10,13)\\ \hline
q=10\_s=990\_2qbf=091\_1 & 10 & 1000 & 0,899 & (10,44)\\ \hline
sqn\_258 & 10 & 29333 & 0,152013091 & (10,42)\\ \hline
sym9\_148 & 10 & 61824 & 0,152173913 & (10,40)\\ \hline
shor\_15 & 11 & 13588 & 0,131586694 & (11,34)\\ \hline
16QBT\_100CYC\_QSE\_1 & 16 & 1776 & 0,18018018 & (16,19)\\ \hline
20QBT\_45CYC\_0D1\_2D2\_0 & 20 & 270 & 0,333333333 & (20,35)\\ \hline
\end{tabular}
\label{tab:benchmarks}
\end{table*}

\end{document}